\documentclass[acmlarge]{acmart}
\setcopyright{acmlicensed}
\acmJournal{TIIS}
\acmYear{2024} \acmVolume{14} \acmNumber{3} \acmArticle{22} \acmMonth{9}\acmDOI{10.1145/3685053}

\AtBeginDocument{%
  \providecommand\BibTeX{{%
    \normalfont B\kern-0.5em{\scshape i\kern-0.25em b}\kern-0.8em\TeX}}}

\newcommand{\system}{MeetMate}

\newcommand{\cl}[1]{#1}

\usepackage{soul}
\usepackage{multirow}
\usepackage{booktabs}
\usepackage{subcaption}
\usepackage{xcolor}

\newcommand{\bs}[1]{\textcolor{blue}{[Bahar: #1]}}

\newcommand{\del}[1]{}
\newcommand{\new}[1]{#1}

\begin{document}

\title[Enabling Interactive Decision Support]{``I Want It That Way'': Enabling Interactive Decision Support Using Large Language Models and Constraint Programming}


\author{Connor Lawless}
\authornote{Work done while at Microsoft.}
\affiliation{%
  \institution{Cornell University}
  \city{Ithaca}
  \state{NY}
  \country{USA}
  \postcode{14850}
}
\email{cal379@cornell.edu}

\author{Jakob Schoeffer}
\authornotemark[1]
\affiliation{%
  \institution{University of Texas at Austin}
  \city{Austin}
  \state{TX}
  \country{USA}}
\email{schoeffer@utexas.edu}

\author{Lindy Le}
\affiliation{%
  \institution{Microsoft}
  \city{Redmond}
  \state{WA}
  \country{USA}}
\email{lindy.le@microsoft.com}

\author{Kael Rowan}
\affiliation{%
  \institution{Microsoft Research}
  \city{Redmond}
  \state{WA}
  \country{USA}}
\email{kael.rowan@microsoft.com}

\author{Shilad Sen}
\authornotemark[1]
\affiliation{%
  \institution{Macalester College}
  \city{St. Paul}
  \state{MN}
  \country{USA}}
\email{ssen@macalester.edu}

\author{Cristina St. Hill}
\affiliation{%
  \institution{Microsoft}
  \city{Redmond}
  \state{WA}
  \country{USA}}
\email{crdaes@microsoft.com}

\author{Jina Suh}
\affiliation{%
  \institution{Microsoft Research}
  \city{Redmond}
  \state{WA}
  \country{USA}}
\email{jinasuh@microsoft.com}

\author{Bahareh Sarrafzadeh}
\affiliation{%
  \institution{Microsoft}
  \city{Redmond}
  \state{WA}
  \country{USA}}
\email{basarraf@microsoft.com}

\renewcommand{\shortauthors}{Lawless et al.}

\begin{abstract}
A critical factor in the success of many decision support systems is the accurate modeling of user preferences. Psychology research has demonstrated that users often develop their preferences during the elicitation process, highlighting the pivotal role of system-user interaction in developing personalized systems. This paper introduces a novel approach, combining Large Language Models (LLMs) with Constraint Programming to facilitate interactive decision support.
We study this hybrid framework through the lens of meeting scheduling, a time-consuming daily activity faced by a multitude of information workers. We conduct three studies to evaluate the novel framework, including a diary study to characterize contextual scheduling preferences, a quantitative evaluation of the system’s performance, and \del{a study of users interacting with a prototypical version of the system.} \new{a user study to elicit insights with a technology probe that encapsulates our framework.}
Our work highlights the potential for a hybrid LLM and optimization approach for iterative preference elicitation, and suggests design considerations for building systems that support human-system collaborative decision-making processes.

\end{abstract}

\ccsdesc[500]{Human-centered computing~Empirical studies in HCI}
\ccsdesc[500]{Computing methodologies~Artificial intelligence}


\keywords{Decision Support, Large Language Models, Constraint Programming, Preference Elicitation, Meeting Scheduling}



\maketitle

\sloppy

\section{Introduction}
Understanding what a user likes and dislikes, that is, their preferences, is often crucial for the success of intelligent systems that provide decision and negotiation support across many domains.
Decision support systems assist the decision-making process itself, helping the user discover and articulate their preferences, understand the link between preferences and decision outcomes, and analyze the steps taken in the process.
However, eliciting preference information from a user to inform such a decision process is challenging.
This challenge manifests not only as a limitation of the algorithms for elicitating or representing preferences but also as an inherent difficulty of users knowing about or expressing their desired solutions. 

A substantial body of literature on judgment and decision-making suggests that preferences are constructed at the time of decision-making rather than being pre-determined or stable~\cite{lichtenstein2006construction,Slovic1995construction}.
Individuals are often unsure of their preferences, hold inconsistent preferences, and change their preferences as they go about making decisions, which can be attributed to the inherent limitations of individuals' memory, attention, and knowledge ~\cite{lichtenstein2006construction,Slovic1995construction, payne1992behavioral, payne1993adaptive}. 
These challenges are shared across widespread decision-making scenarios, such as meeting scheduling, travel planning or purchasing a house.
Preference elicitation tools and techniques \cite{goldsmith2008preference} are used to assist human decision-makers in overcoming some of their inherent memory or knowledge limitations, extracting information from the user's mental representation of that preference and translating it into a representation through which the system can reason.
However, the constructive nature of human preferences implies that the interaction between the system and a user can greatly influence the quality of the preference information \cite{chen2009interaction, pommeranz2012designing, johnson2005making} and the user's acceptance of the results provided by the system \cite{carenini2002constructed, bettman1998constructive}.
Prior work has shown that such preference elicitation systems need to be (1) \textit{collaborative} to facilitate seamless information exchange between the user and the system~\cite{peintner2008preferences}, (2) \textit{contextualized} to the decision-making process \cite{payne1993adaptive, Slovic1995construction, pu2003user, carenini2002constructed}, and (3) \textit{incremental} to account for evolving preferences \cite{pu2003user, pommeranz2012designing, ardissono2003framework}.
%
In building a preference elicitation system, Large Language Models (LLMs) provide a promising opportunity to support the naturalistic and contextual elicitation of user preferences via chat. 
Recent work has demonstrated their potential to work on collaborative tasks \cite{lin2023decision}, and its integration into chat enables a system that can be responsive to both the context of the decision-making task and the incremental evolution of preferences.


Even once preferences are elicited, finding a solution that best meets them represents a challenging decision-making task.
Optimization techniques such as Constraint Programming (CP) are a natural fit for solving this complex reasoning problem and have been successfully applied in a range of multi-attribute decision problems.
\del{CP specifically aligns with multi-attribute decision theory [31, 35, 41], where preferences are represented as utility functions or embedded as constraints, and constraint satisfaction techniques are leveraged to manage trade-offs [78].}
One major shortcoming of traditional optimization approaches is that they require full knowledge of the model's specification---including a fully specified objective and all relevant constraints in a structured format.
This is at odds with how users approach complex and contextual decision-making tasks where they often do not know relevant preferences unless presented with candidate solutions.
Furthermore, customizing optimization models requires expertise in coding and optimization, an unrealistic barrier for most users.
Here, in addition to their potential in eliciting preferences, LLMs provide another opportunity to overcome this technical hurdle by translating natural language into structurued forms that can be used by an underlying optimization solver.
Finally, these optimization approaches must function as dynamic and incremental reasoning engines: users actively construct and adjust their preferences while the system continuously learns and embeds their preferences, offering increasingly improved suggestions until a satisfactory one is achieved. 

In this paper, we explore a framework that supports  iterative preference elicitation, aligning with how humans develop or construct preferences in their decision-making.
Elicited preferences are embedded into specific constraints that can be solved within an optimization setting.
We leverage (1) LLMs to facilitate natural and intuitive communication of preferences and suggestions as well as to reason about how preferences should be incorporated and translated into structured constraint functions, and (2) CP techniques to reason over the space of outcomes and find optimal solutions given incorporated constraints. 

We study this framework for interactive decision support through the lens of meeting scheduling---a routine, tedious, and time-consuming daily activity faced by a multitude of information workers.
At its core, scheduling a meeting is a complex spatio-temporal task that is highly contextual, and finding a suitable meeting time may require reasoning over conflicting preferences, constraints, and priorities \cite{baharcscw}.
It is also an incremental and continuous process that requires iteratively refining candidate solutions in a more or less greedy fashion \cite{brzozowski2006grouptime}. 
Therefore, interactive decision support tools are a natural fit for scheduling as they can provide contextualized and interactive support for the complex reasoning that is necessary to complete the task. \cl{We specifically study professional meeting scheduling for information workers. This setting comes with a host of implicit scheduling constraints including working hours, implicit power dynamics (i.e., for managers vs. team members), and imperfect calendar information (i.e., undisclosed personal scheduling constraints that are not reflected in a work calendar) that provide a rich setting in which to study contextual preference elicitation.}
\new{Moreover, our work focuses on the process of a meeting organizer finding a candidate meeting time.
This is in line with most modern industry-scale tools for meeting scheduling, such as Microsoft Outlook, where attendees' information is limited to their calendar's free/busy time. Previous work has also shown that organizers typically have more power \new{than other participants} over where a meeting lands  \cite{mok2023challenging}.
We fully acknowledge that meeting scheduling is a distributed negotiation task amongst multiple participants and encourage follow-up research that builds on our organizer-centric investigation.}

\begin{figure}[!t]
     \centering
      \includegraphics[width=\textwidth]{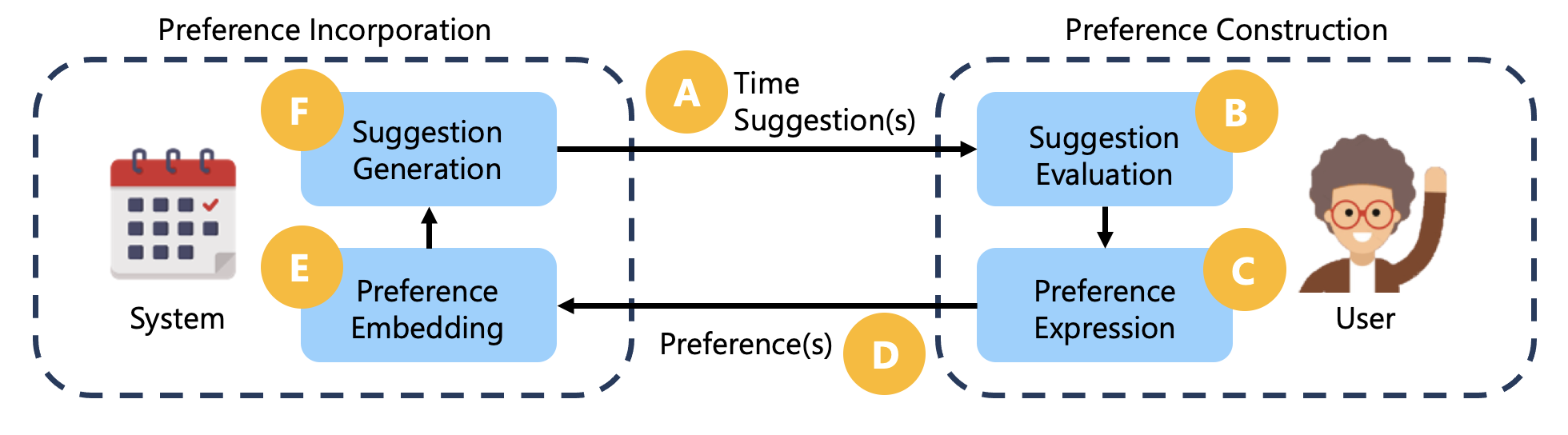}
         \vspace{-0.4cm}
           \caption{An overview of the interactive loop for conversational decision support. The loop is initiated via an initial time suggestion (or set of suggestions) by the system presented to the user (A). During \textit{Preference Construction} users evaluate a proposed suggestion (C) and either express a new preference (D) to improve the given suggestion, or accept the current suggestion. New preferences are integrated into the system during \textit{Preference Incorporation} which requires the system to both embed the stated preference into the system (E) and then use it to generate a new time suggestion (F). This process is iterated until a suitable time is found.\label{fig:interaction_loop}
           }
    \Description{Two components: Preference Construction and Preference Incorporation. Preference Construction is driven by the User and consists of (B) Suggestion Evaluation followed by (C) Preference Expression, which passes a (D) Preference to the System. Preference Incorporation is driven by the System and contains (E) Preference Embedding followed by (F) Suggestion Generation, which passes a (A) Time Suggestion to the User.}
\end{figure}

We introduce \system, a hybrid LLM and CP approach that capitalizes on the benefits of both technologies to build an interactive decision support system for meeting scheduling. 
\system~ is designed to facilitate human-system collaboration that encompasses two key components: Preference Construction and Preference Incorporation. \textit{Preference Construction} is the process by which users of the system \textit{evaluate} candidate decisions (i.e., suggested meeting times) and \textit{express} new preferences and constraints that need to be integrated into the system.
\textit{Preference Incorporation} is a mirror of the construction process from the system perspective and involves \textit{embedding} expressed user preferences in a form that can be used by the system to \textit{generate} new suggested decisions.
We refer to the entire process of iteratively eliciting preferences via generating and critiquing suggestions as \textit{Preference Elicitation}. 
Because eliciting preferences is an iterative process, \system~ unifies Preference Construction and Preference Incorporation into one dynamic loop where preferences are elicited in conjunction with refined suggestion generations.
Figure \ref{fig:interaction_loop} outlines the key components of our interactive decision support system for meeting scheduling. 

We conducted three studies to understand preference elicitation in the context of meeting scheduling.
First, we study how preferences are constructed via a diary study that collected naturalistic preferences in the moment of scheduling (Section \ref{sec:diary}). 
Informed by the results of the diary study, we then introduce \system~(Section \ref{sec:architecture}) and quantitatively evaluate the system's capabilities for preference incorporation (Section \ref{sec:quant_eval}). Our results show that LLMs can generate high-quality functional representations of scheduling constraints that achieve high precision and recall when compared to human expert implementations of the constraints.
Finally, we \new{use \system~ as a technology probe to} conduct a user study (Section \ref{sec:user_study}) \del{with a prototype system}\new{,} to observe the iterative preference elicitation flow in situ\new{,} and to inform future design improvements for intelligent interactive decision support systems.
Through our study, we confirm that participants' preferences are indeed constructed through the iterative preference elicitation process of evaluating dynamic time suggestions that incorporate their feedback.
We also identify that system explanations play an important role in bridging the gap between understanding the system and expressing preferences, and contrasted suggestions not only give users a sense of agency but also help make the decision-making process efficient.
Our work highlights the potential for a hybrid LLM and optimization approach for iterative preference elicitation and design considerations for building systems that support human-system collaborative decision-making processes.

\section{Background and Related Work}
Our paper connects to a long line of work on both human preferences, and approaches by which to elicit them.
From a methodological perspective, we build upon recent work at the intersection of large language models, mathematical optimization, and mechanisms by which they interact. 

\label{sec:RW}
% ...

\subsection{Human Preferences and Design Implications}
\label{subsec:HumanPreferences}

In 2002, \citet{carenini2002constructed} described a shift in classical decision theory towards constructive preferences, contrary to the prevailing view of stable and rational preferences (e.g., \cite{doyle2004prospects}).
Psychological studies have shown that preferences are indeed often constructed rather than stable, and individuals may hold inconsistent or evolving preferences~\citep{Slovic1995construction,lichtenstein2006construction, johnson2005making}.
This constructive nature of preferences has been emphasized, suggesting that preferences are more like architectural creations than archaeological discoveries \cite{gregory1993valuing, payne2000measuring}.
In various decision-making theories \cite{janis1977decision, mills1971anticipated, montgomery1999decision, russo1996distortion}, it has been posited that people construct preferences when faced with decisional conflicts.
Preferences are not well-defined in most situations but are constructed in the context of decision-making, particularly when conflicts arise.
This process allows individuals to adjust their initial preferences to make confident decisions, especially when one option clearly outweighs others~\citep{simon2008transience}.
Incremental preference elicitation has received attention \cite{pu2003user, pommeranz2012designing, ardissono2003framework} due to the uncertain nature of user goals and preferences.
It involves improving a user's preference model over time as they interact with a system.

Preference construction has also been explored within the framework of multi-attribute utility theory (MAUT)~\citep{edwards1992multiattribute,keeney1993decisions, hammond2015smart}, where preferences are represented as utility functions.
Research in this area has shown that people tend to increase their preferences for attributes of chosen options while decreasing preferences for attributes of rejected options.
Similar effects have been observed in negotiation settings~\citep{curhan2004dynamic}.
These findings align with the goal of maximizing the ease of justifying a decision, as identified by \citet{bettman1998constructive}.

To develop an effective decision support system, designers must take into account the process of human preference development \cite{pu2003user, pommeranz2012designing, chen2004survey}.
Relevant literature provides key insights for designing effective decision support systems in two main areas.
First, research in human-computer interaction (HCI) and user-centered design highlights the paramount significance of interactivity and collaboration within decision support systems.
Effective decision support necessitates a collaborative environment wherein both the user and the system can seamlessly exchange information and jointly tackle decision-making challenges.
Second, studies in supporting preference construction offer further guidance on structuring this collaborative process.
The collaboration involves two critical elements: (1) natural interaction, that is, the system user interface (UI) should offer user-friendly tools for users to express, revise, and provide feedback on their preferences and constraints, as well as critique system suggestions; (2) iterative and transparent system recommendations, that is, the system should contextualize the decision-making, enabling users to understand decision consequences  \cite{payne1993adaptive, Slovic1995construction, pu2003user}.
Providing concrete examples for user critique \cite{pu2003user, chen2012critiquing, mcginty2010evolution}, along with immediate feedback and explanations, helps users grasp the correlation between preferences, constraints, and proposed solutions \cite{carenini2002constructed}. 


Our work primarily centers on the information flow from the system to the user, aligning with the process of human preference development and decision-making.
It underscores the significance of contextualized elicitation mechanisms, concrete examples for user critique, immediate feedback with explanations, incremental suggestion generation, and the utilization of constraint satisfaction as part of the reasoning engine.

\subsection{Preference Elicitation Techniques}
\del{To create a user-friendly system, designers must consider the development of human preferences, and as such,}
Preference elicitation \new{has been studied} in artificial intelligence (AI), recommender systems, and HCI~\citep{goldsmith2008preference}. Various techniques exist for inferring user preferences, including rankings, corrections, critiques, ratings, queries, and past behavior.
Preference elicitation methods can be broadly categorized into two types: \textit{implicit} and \textit{explicit}. Implicit methods involve learning user preferences from behavior, aligning with Revealed Preference Theory~\citep{richter1966revealed}. In contrast, explicit methods, such as collecting ratings, actively involve users in constructing and reflecting on their preferences. Our approach falls into the explicit category, allowing users to specify their preferences during the decision-making process.
Our method draws inspiration from techniques like \textit{Similarity and Tweaking}~\citep{carenini2003towards,jannach2021survey}, and \textit{Example-Critiquing Interaction}~\citep{chen2012critiquing}.
These approaches involve users in the recommendation process by having them specify attributes or critique candidate options. Similarly, our system follows an iterative approach, enabling users to critique time recommendations within a chat interface.

Our work is also related to preference elicitation in decision support systems, where interactive tools assist users in decision-making.
Many decision support systems rely on MAUT~\citep{keeney1993decisions} to represent preferences as utility functions.
However, constructing these utility functions in collaboration with users is challenging. 
Traditional methods like absolute measurement and pairwise comparison (active elicitation)~\citep{aloysius2006user} may lead to misalignment between users' qualitative mental models of preferences and the system's quantitative representation.
\del{In the context of calendar scheduling, which is the primary context of this work, understanding user preferences is essential.
Herein, previous research has used various approaches, including machine learning on existing calendars [61], showing users sample schedules [16], and allowing direct input of parameters [40].
Our approach stands out by allowing users to specify new preferences flexibly, accommodating more diverse contextual scheduling preferences.}

\subsection{User Scheduling Preferences}
Whether a user is arranging a new meeting or responding to a request, they consider multiple objectives, such as minimizing disruptions, travel time, and preferred locations \cite{berry2007balancing,oh2005calendar}.
An effective calendar scheduling system needs to understand these user preferences to provide useful recommendations \cite{oh2005calendar}, which has inspired a line of work on eliciting and incorporating user preferences into calendar scheduling systems \cite{dent1992personal,mitchell1994experience,sen1997satisfying,chajewska2000making,tullio2002augmenting,oh2005calendar,gervasio2005active,brzozowski2006grouptime,krzywicki2010adaptive,kim2018learning}.
Prior work has explored complex and often contradictory preferences in the context of meeting scheduling (see \cite{baharcscw} for a broader review).
\del{, which highlight the importance of understanding user preferences in an effective calendar scheduling system [66].
However,}
Understanding user scheduling preferences is a complicated task that suffers from many of the same challenges as the broader preference elicitation setting.
Users may not be fully aware of their preferences, and may find the elicitation process itself burdensome.
Attempting to address the burden of explicit preference elicitation, prior work has used machine learning on user's existing calendar to learn scheduling preferences \cite{mitchell1994experience, brzozowski2006grouptime, gervasio2005active}. 
There have also been works focused on explicit preference elicitation by showing users sample schedules \cite{brzozowski2006grouptime, viappiani2006evaluating}, or allowing them to directly input parameters into a fixed preference model \cite{haynes1997automated}. 
\citet{berry2007balancing} took a user-centered approach, capturing criteria through structured interviews and using an interactive interface.
However, all these approaches require a fixed preference model that users, either implicitly or explicitly, populate with their own preferences.
In contrast, our approach allows users to specify new preferences that may not be captured in a fixed preference model (e.g., realizing they want a 15 minute break only after seeing more than 2 hours of back-to-back meetings).
This flexibility better aligns with the diversity of contextual scheduling preferences we captured in our diary study (Section~\ref{sec:diary}).

Given the routine, tediuous, and time-consuming nature of meeting scheduling, a wide array of tools have been developed to help users schedule meetings \cite{brzozowski2006grouptime, cranshaw2017calendar, kim2018learning, oh2005calendar,berry2011ptime}.
However, many existing calendar tools are too rigid to accommodate the diversity, \new{uncertainty, and evolution} of user scheduling preferences.\del{---requiring users to explicitly search over candidate times, evaluating times against their own personal preferences, until they find a satisfactory slot.}
Recent work has tried to automate the task of finding a candidate meeting time via machine learning \cite{brzozowski2006grouptime}, but gives users no recourse to customize or tweak suggested times if the ones provided do not work.
Other tools, such as Doodle and When2meet, have focused on the group nature of meeting scheduling, allowing users to share their availability and find a shared meeting time.
However, these tools typically only allow users to give binary feedback (i.e., whether or not they are available for a given time slot) and do not allow users to negotiate over subjective preferences and their relative importances.
For instance, users may not want to put morning availability for a meeting unless it is the only option for which all participants are available.
In contrast, our proposed system allows users to specify scheduling preferences and their relative importance in natural language and can accommodate more diverse preferences then existing systems. 

\subsection{LLMs and Optimization}


Large Language Models (LLMs) have emerged as a promising tool for helping accomplish a number of tasks such as generating code \cite{chen2021evaluating}, writing \cite{yuan2022wordcraft}, and visualizing data \cite{dibia2023lida}. LLMs have also been applied to convert human preferences directly into automated decision-making systems~\cite{li2023eliciting} or to solve constraint satisfaction problems~\cite{yuksekgonul2023attention,abdin2023kitab}.
However, they are well known to suffer from hallucinations, seemingly plausible text predictions that are in fact non-factual \cite{yuksekgonul2023attention,welleck2019neural}, and poor mathematical reasoning ability \cite{qian2022limitations}.
Recent work has aimed to address these shortcomings by helping LLMs decompose complex tasks into simpler steps (e.g., \cite{wei2022chain, yao2022react,yao2023tree}) and augmenting them with external tools (e.g., \cite{schick2023toolformer, thoppilan2022lamda, nakano2021webgpt}), such as a code interpreter.
We refer to \citet{mialon2023augmented} for a survey of other related approaches.
Our work builds upon this line of work by decomposing the complex problem of suggesting meeting times into structured sub-components (detailed in Section \ref{sec:architecture}).
We also employ an optimization solver as an external tool to perform complex reasoning over meeting times. 
This is in contrast to other existing work on collaborative decision-making between LLMs and humans, such as by \citet{lin2023decision}, which requires the LLM to generate the solutions.

Mathematical optimization is a powerful tool used in various industrial applications~\cite{bertsimas1997introduction}.
In this paper, our focus is on (Weighted) Constraint Programming~\citep{rossi1999constraint}, a modeling paradigm where the optimization problem involves a set of constraints with different weights.
The objective is to find a solution with the optimal score, determined by the total weight of satisfied constraints.
Traditionally, constraint programming demands a complete specification of the optimization model, including all constraints and their coefficients.
This poses challenges for preference elicitation, where users' constraints and preferences are often unknown and expressed within the context of proposed solutions\new{---as discussed earlier.}
Recent research has addressed this by tackling constraint programming problems with incomplete information, requiring the elicitation of certain data and preferences~\citep{tabakhi2017preference,tabakhi2017preference2,tabakhi2021preference,tabakhi2022incomplete,xiao2020embedding}.
However, existing approaches primarily focus on the algorithmic aspects of what information to elicit, rather than the mechanism for eliciting it.
This distinction is crucial, particularly in settings like meeting scheduling, where non-expert users may lack the modeling expertise to formulate the constraints they want in the model.

Natural language processing holds promise in simplifying interactions with optimization models for non-expert users.
Recent work has employed LLMs for users to query models in areas like supply chain management~\citep{li2023large} and identifying infeasible subsystems~\citep{chen2023diagnosing}.
However, these chat-based systems primarily aid in understanding existing models, not constructing new ones.
To address the challenge of formulating optimization problems, recent research \new{has focused} on translating natural language descriptions into formal optimization models. \citet{ramamonjison2022augmenting} initiated this effort, leading to the NL4Opt competition at the NeurIPS conference and related work on named entity recognition and semantic parsing~\citep{ning2023novel,tsouros2023holy,dakle2023ner4opt,prasath2023synthesis}.
These approaches typically address static contexts, translating a single text paragraph into an optimization model.
In contrast, our study focuses on an interactive model, capturing user preferences during optimization model construction.
We emphasize translating natural language into Python constraints within a constraint programming framework, distinct from traditional named entity recognition for linear and integer programming. 

\section{Preference Construction: Understanding Contextual Scheduling Preferences}
\label{sec:diary}
We first investigated preference construction in the context of meeting scheduling. Specifically, we aimed to identify and characterize the types of contextualized preferences elicited in response to given time suggestions. In the absence of an existing interactive system through which we could analyze users preferences in response to \textit{dynamic time suggestions}, we conducted a diary study at a large technology company that elicited user preferences in response to \textit{static} time suggestions from an existing calendar tool. The findings of our diary study were used to inform the design of the system (Section~\ref{sec:architecture}) and create a new quantitative benchmark for its performance (Section~\ref{sec:quant_eval}). We evaluate the system in the context of \textit{dynamic} time suggestions in the user study outlined in Section \ref{sec:user_study}.

\subsection{Study Protocol}
To capture contextual scheduling preferences for a given meeting, participants were asked to complete a diary entry for each meeting they scheduled during a one-week study period. Participants were instructed to first phrase their initial meeting scheduling request to a pretend chat-based interface (e.g., ``I want to schedule a 30 minute 1:1 with my manager within the next week"), and then enter the needed information into a widely used commercial calendaring tool that generates time suggestions for a meeting. In response to the suggested times, participants were then asked to express any preferences up to 5 times (e.g., ``I can't meet before 10am so that I can drop my kids off before work"). They were also asked to reflect on whether there was information missing from their calendar that could have captured their stated preference. 
Finally, participants were also asked to include information about the meeting they were scheduling including the number of attendees, its duration, and the time horizon in which it was to be scheduled. 
We recruited 64 adult participants based in the United States through an e-mail advertisement at a large technology company who scheduled their own meetings (i.e., did not have an administrative assistant) and had at least 2 meetings per week. 
Participants self-reported as working in the roles of a manager (39\%), program manager (30\%), developer (17\%), and researcher (9\%). 39\% of participants reported attending under 20 meetings per week, 37\% attended between 20-30, and the remainder reported over 30 meetings per week. We collected diary entries for 114 unique meetings with averages of 3.95 attendees and 35 minutes. 58\% of the meetings were expected to be fully virtual, 20\% hybrid, and 4\% fully in-person (the modality of the remainder of meetings was uncertain at the time of scheduling). \cl{Despite no such requirement in the study instructions, all meetings scheduled during the diary study pertained to work scheduling.}
In total, we collected 114 instances of the initial phrasing of the scheduling request, 211 contextual preference expressions, and 197 missing information descriptions. \cl{We donated \$1 dollar to charity for each diary entry completed, totalling to \$114. Participants were informed of this reward scheme and the selected charity at the time of study enrollment.}
The study design and protocol were approved by the leading institution's Institutional Review Board (IRB).

\subsection{Analysis \& Findings}
We qualitatively analyzed all preference and missing information entries following ~\cite{charmaz2006constructing}.  
Researchers conducted open-coding of 30\% of the entries. Initial codes were iterated and grouped into coarser categories of preferences and missing information to develop a codebook which was subsequently used to conduct closed-coding of the remainder of the data.
While our survey asked users to express preferences in response to suggested times, we also observed users requesting additional actions such as querying calendar data (e.g., ``Are [external participants] available during this time?"), drafting an agenda (e.g., ``Draft the agenda based on previous meeting with the person"), or sending messages (e.g., ``send a message asking to move the meeting earlier rather than later"). We restricted our analysis to focus only on preferences, as opposed to these auxiliary actions. 
We also labeled each of the initial phrasings of the scheduling requests for the presence of the duration of the meeting, the attendees, and the expected time frame. 
We then reviewed the labeled and coded entries to highlight four key takeaways from our analysis.

 \begin{figure}[!t]
     \centering
     \begin{subfigure}[b]{0.49\textwidth}
         \centering
         \includegraphics[width=\textwidth]{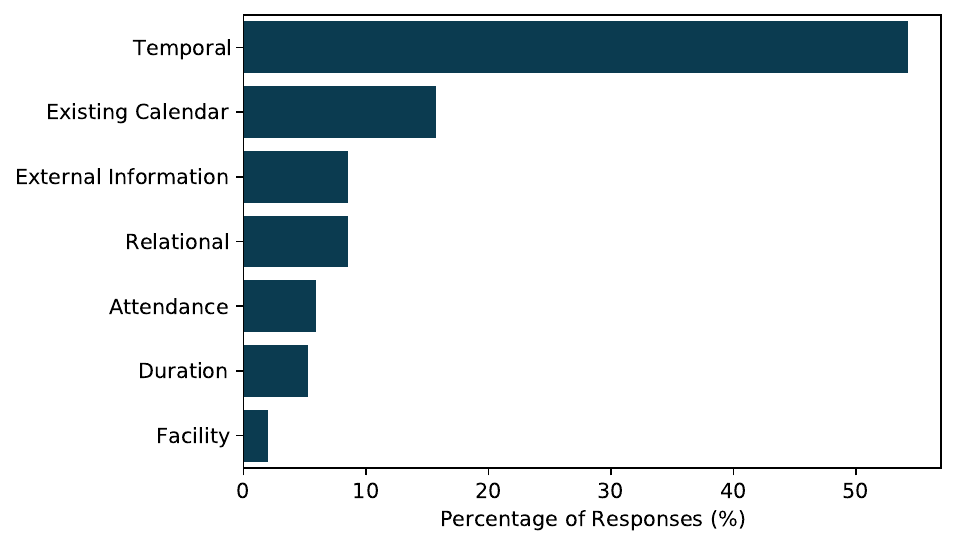}
         \caption{Contextual Preferences}
         \label{fig:pref_breakdown}
     \end{subfigure}
     \begin{subfigure}[b]{0.49\textwidth}
         \centering
         \includegraphics[width=\textwidth]{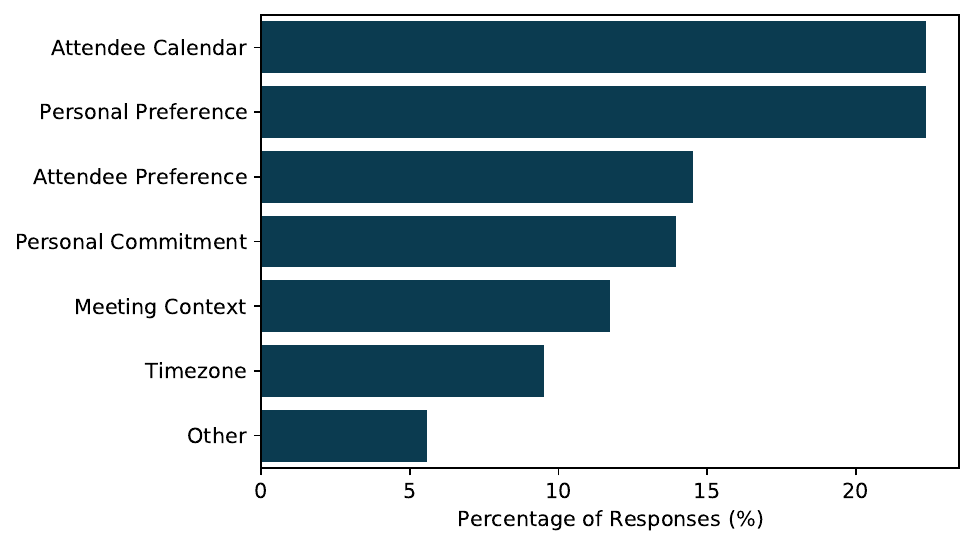}
         \caption{Missing Information}
         \label{fig:missing_info_breakdown}
     \end{subfigure}
        \caption{Breakdown of the percentage of diary study responses for (a) categories of contextual scheduling preferences and constraints, and (b) information missing from the calendar tool that inform the preference or constraint.}        
\Description{Two horizontal bar graphs, titled (a) Contextual Preferences and (b) Missing Information. Contextual Preferences shows scheduling preference category on the Y axis against percentage of responses on the X axis. Seven bars shown, ordered from largest to smallest category: Temporal- 54\%, Existing Calendar- 16\%, External Information- 8.5\%, Relational- 8.5\%, Attendance- 6\%, Duration- 5\%, and Facility- 2\%. Missing Information shows missing information type on the Y axis against percentage of responses on the X axis. Seven bars shown, ordered from largest to smallest information type: Attendee Calendar- 22\%, Personal Preference- 22\%, Attendee Preference- 15\%, Personal Commitment- 14\%, Meeting Context- 12\%, Timezone- 9\%, and Other- 6\%.}
\end{figure}

\subsubsection{Scheduling preferences are diverse} In our analysis of the stated preferences, we identified 7 key types of contextual preferences for scheduling a meeting:
\begin{itemize}
\item \textbf{Temporal: } Preferences for the time of day or day a week a meeting is scheduled (e.g., ``I prefer Tuesday or Wed next week because we need a day of prep (P9)'').
\item \textbf{Existing Calendar: } Preferences related to how to handle existing items on a user's calendar (e.g., ``It's ok to propose times during the learning day all day event (P138)'' .
\item \textbf{External Information: } Preferences related to events external of the user's work calendar (e.g., ``I need to meet before I pick my child up (P52)'').
\item \textbf{Relational: } Preferences for how the new meeting is scheduled in relation to existing items on the calendar (e.g., ``I need at least a 5 minute gap between meetings for a break (P1)'').
\item \textbf{Attendance: } Preference over the relative importance of different attendees availability for the meeting (e.g., ``Both [co-worker 1] and [co-worker 2] are required to be available at this meeting (P27)'').
\item \textbf{Duration: } Preferences related to the duration of the meeting (e.g., ``I need this working session to be 1 hour long (P40)'').
\item \textbf{Facility: } Preferences related to physical meeting space required for the meeting (e.g., ``I need a meeting room in my building that is free and can accommodate all in-person attendees (P70)'').
\end{itemize}

Figure \ref{fig:pref_breakdown} shows a breakdown of the expressed preferences categorized along these types. We found that a majority (54\%) of the elicited preferences are temporal, highlighting that providing a way to adjust suggested times to simple time and day requests could satisfy a large fraction of users' unmet scheduling preferences. However, even within temporal preferences, participants exhibited a wide range of preferences in terms of both preferred meeting time and how the preferences were phrased. 
This underscores the limitations of graphical interfaces: the wide scope of user preferences makes it difficult to fully capture all possible preferences within a form or pre-fixed set of options. Preference elicitation systems need to be able to flexibly capture diverse user preferences via mediums such as natural language in chat.

\subsubsection{Scheduling requests are vague} When scheduling a meeting, many existing time suggestion features require a list of attendees, a duration, and a time frame (e.g., the next two weeks). However, we found that the initial scheduling requests in our data did not include the duration of the meeting 57\% of the time, the expected time frame 28\% of the time, and even the attendees of the meeting 4\% of the time. This highlights a key challenge in performing scheduling via chat. Chat-based scheduling tools need to elicit key pieces of information from the user before generating an initial time suggestion or provide a means for users to validate assumptions generated by the system (e.g., default durations for meetings).

\subsubsection{Calendars are incomplete} A key finding of the diary study is that users' calendars rarely reflect the complexities of their true scheduling preferences and considerations. In our analysis of missing information entries, we identified 6 types of missing information: 
\begin{itemize}
	\item \textbf{Attendee Calendar}: Information on attendees' availability including both existing events and their relative importance (e.g., ``Whether the meetings on their calendar are actual meetings or self-imposed work blocks (P122)'').
	\item \textbf{Personal Preference}: Information on user's personal scheduling preferences (e.g., ``My personal preference is that meeting align into up to 2 hour blocks, not be spread throughout the day breaking up the rest of my focus time. this is implicit knowledge about my preferences (P38)'').
	\item \textbf{Attendee Preference}: Inferred preferences about the attendees of the meeting (e.g., ``Coworker's working hours shown in her calendar does not accurately reflect the actual working hours (P34)'').
	\item \textbf{Personal Commitment}: Scheduled events or obligations that were missing from the user's work calendar (e.g., ``I was planning to be oof\footnote{\cl{OOF refers to being `out of office`.}} but didn't make it in the calendar (P70)'').
	\item \textbf{Meeting Context}: Details about the meeting to be scheduled that inform how the meeting should be scheduled (e.g., ``I have a 1:1 with one of the attendees at 2 so i'd like to have the meeting earlier so we can focus on other topics at the 1:1 (P86)'').
	\item \textbf{Timezone}: Information about the timezone of various participants and what an appropriate meeting time is for their working hours (e.g., ``My colleague is on the east coast (P66)'').
\end{itemize}

Figure \ref{fig:missing_info_breakdown} shows a breakdown of the missing information expressed during the diary study along these axes. While an ideal calendar system may be able to internalize some of this missing information (i.e., by asking users to input all their personal commitments), the scope and diversity of the missing information makes capturing it all in existing systems near impossible. This highlights the importance of a flexible interface through which users can inform the scheduling system of context and information missing from a user's calendar. 

\subsubsection{Users value temporal diversity} Our analysis of the preference expressions also revealed that participants desire multiple diverse time suggestions to help speed up the scheduling process. For instance, P3 mentioned that suggestions that are ``back to back on the same day is not helpful.'' 
P99 wanted ``time slots for 3 days'' to ``make scheduling much faster.'' Notably in these responses, participants characterized diversity with respect to the time and day of the meeting, underscoring the importance of \textit{temporal} diversity in meeting time suggestions.



\section{Preference Incorporation: \system~System} \label{sec:architecture}

We now introduce a novel framework for performing preference incorporation in the context of meeting scheduling. Given the diversity of user preferences discovered in the diary study, such a system needs to be able to flexibly incorporate a wide range of preferences. Furthermore, prior work in preference elicitation has highlighted the importance of such a system being both contextual and interactive. 

Towards a flexible, contextual, and interactive decision support system, we introduce \system, a chat-based meeting scheduling tool built in Python that allows users to interactively refine meeting time suggestions via natural language requests.
A key component of our approach is an underlying constraint programming optimization model to generate new time suggestions (detailed in Section \ref{sec:cp_solver}), which performs the complex mathematical reasoning required to find suitable times. We model each scheduling preference as a function, called a constraint, that checks whether a candidate meeting time meets the preference (e.g., ``Not scheduled before 2pm"). Each constraint also has an associated weight. Intuitively, higher weights correspond to constraints that are more important. Instead of requiring users to specify an exact numerical weight for each preference, the system instead keeps an underlying \cl{ordered} list of the constraints which are then translated into weights such that constraints with higher priority \cl{(i.e., earlier in the list)} are always satisfied before those of lower priority \cl{(i.e., later in the list)}. Each constraint is formalized as a Python function that is used by a solver to generate new time suggestions. To embed natural language scheduling preferences into Python functions, we leverage LLMs to translate user chat messages to operations on the weighted list of scheduling constraints. This hybrid framework capitalizes on the benefits of both LLMs and Optimization, using LLMs to flexibly embed naturalistic scheduling constraints into Python functions and using optimization to generate new time suggestions that are responsive to the embedded preferences. This occurs within an interactive chat environment which allows the system to be both contextualized to a given scheduling instance and interactive.  

\begin{figure}[!t]
     \centering
      \includegraphics[width=\textwidth]{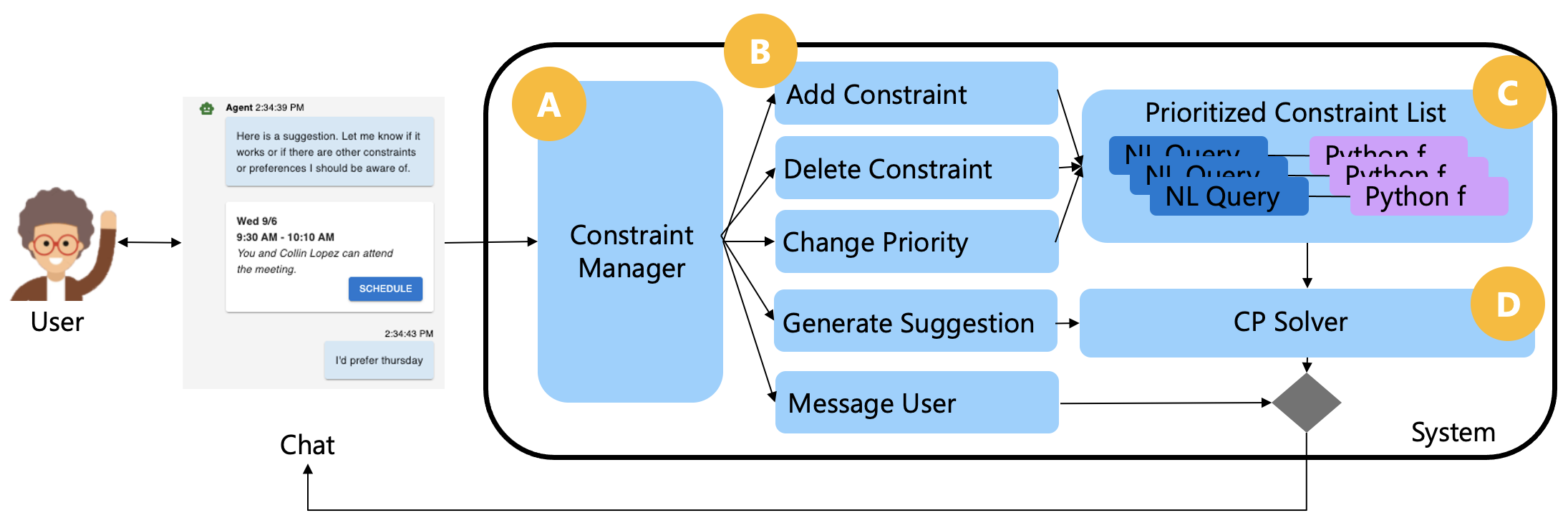}
         \vspace{-0.4cm}
        \caption{Overview of \system~ system architecture. Chat messages from users are translated into actions by the Constraint Manager Component (A), which selects from 5 actions including the Add Constraint Action (B) which translates natural language scheduling preferences into python functions. The system maintains an ordered list of scheduling constraints with priorities (C) that are used by the Constraint Programming Solver (D) to generate new time suggestions.  \label{fig:system_arch}}        
\Description{Three components: User, Chat, and System. The User sends messages via Chat to the System, which returns a response via Chat to the User. The System has four subcomponents: (A) Constraint Manager, (B) System Operations, (C) Prioritized Constraint List, and (D) Constraint Programming Solver. Messages enter the System via the Constraint Manager, feeding into System Operations, which includes Add Constraint, Delete Constraint, Change Priority, Generate Suggestion, and Message User. Add Constraint, Delete Constraint, and Change Priority feed into Prioritized Constraint List, which includes Natural Language Queries mapped to Python Functions. Both Generate Suggestion and Prioritized Constraint List feed into Constraint Programming Solver. Both Message User and Constraint Programming Solver feed into the System’s response back to the User.}
\end{figure}

When a user begins to schedule a meeting, they must first enter the list of attendees, the duration, and the time horizon for the meeting into a form. Using a form ensures that users include the minimum viable amount of information to schedule a meeting, and alleviates the vague scheduling behavior we discovered during the diary study. \cl{This aligns with many existing calendar UIs which require users to specify a duration and attendees before generating candidate meeting times.} \cl{Similar to popular institutional scheduling tools like Microsoft Outlook and Google Calendar, the system is given access to each attendee's public free/busy times, but is not given information about the meetings themselves (e.g., names, other attendees) to mitigate privacy concerns.} Once a user enters the information, the system generates an initial time suggestion that maximizes the availability of the desired attendees. The system returns the time suggestion and initiates a chat so that users can interact with the system to refine the suggested meeting time. For every new chat message entered in the chat, the \textit{Constraint Management} Component (Section \ref{sec:const_mgt}) translates the request into an action for the system to take. One such action is the addition of a new constraint (see Section \ref{sec:const_gen}), which requires translating a natural language preference into a Python function. These Python functions are then used within a constraint programming solver (Section \ref{sec:cp_solver}) to generate new time suggestions. An overview of the system is included in Figure \ref{fig:system_arch}.

\subsection{Constraint Management} \label{sec:const_mgt}
\begin{figure}[ht]
     \centering
      \includegraphics[width=\textwidth]{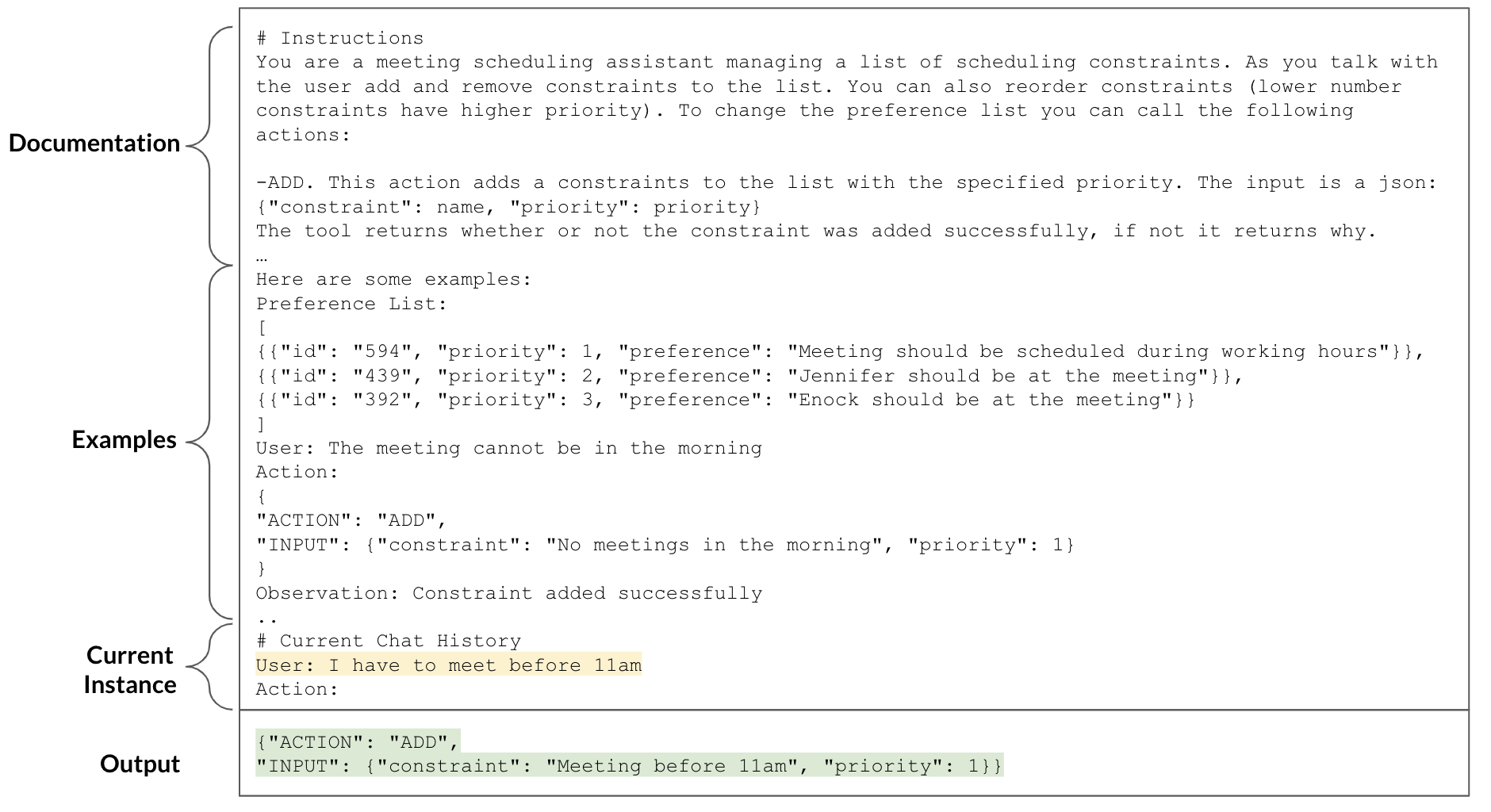}
        \caption{Sample prompt for the constraint management component. Orange highlight reflects scheduling instance-specific inputs. Green highlights the output of the LLM. \label{fig:manager_prompt}}
        \Description{Three input components: Documentation, Examples, and Current Instance as well as an Output component. In this sample, Documentation includes: ‘# Instructions. You are a meeting scheduling assistant managing a list of scheduling constraints. As you talk with the user add and remove constraints to the list. You can also reorder constraints (lower number constraints have higher priority). To change the preference list you can call the following actions: -ADD. This action adds a constraints to the list with the specified priority. The input is a json: {"constraint": name, "priority": priority}. The tool returns whether or not the constraint was added successfully, if not it returns why…’ Examples include: ‘Here are some examples: Preference List: [{{"id": "594", "priority": 1, "preference": "Meeting should be scheduled during working hours"}}, {{"id": "439", "priority": 2, "preference": "Jennifer should be at the meeting"}}, {{"id": "392", "priority": 3, "preference": "Enock should be at the meeting"}}]. User: The meeting cannot be in the morning. Action: {"ACTION": "ADD", "INPUT": {"constraint": "No meetings in the morning", "priority": 1}}. Observation: Constraint added successfully…’ Current Instance includes: ‘# Current Chat History. User: I have to meet before 11am. Action:’ Output includes: ‘{"ACTION": "ADD", "INPUT": {"constraint": "Meeting before 11am", "priority": 1}}.’ From Current Instance, the line ‘User: I have to meet before 11am’ is highlighted in orange. The entire Output is highlighted in green.}
\end{figure}

When a user enters a new chat message, an LLM-powered \textit{Constraint Management} component is prompted (see Figure \ref{fig:manager_prompt}) to select one of five actions to take: 

\begin{itemize}
\item \textbf{Add Constraint}: This action generates a new scheduling constraint and priority, and calls the constraint generation functionality outlined in Section \ref{sec:const_gen}.
\item \textbf{Change Priority}: Change the priority of a specified constraint.	
\item \textbf{Delete Constraint}: Removes a given scheduling constraint.
\item \textbf{Message User}: Sends a message back to the user.
\item \textbf{Generate Suggestion}: Calls the CP solver to generate a new time suggestion (detailed in Section \ref{sec:cp_solver}), and returns it to the user.
\end{itemize}
 
Notably, the constraint management component is given as input the entire chat history between the user and the agent, and the current list of scheduling constraints. This allows the actions to be informed by not only the most recent chat message but the context in which the message is taking place. For instance, if a user tries out a scheduling constraint (e.g., ``How about meeting on thursday?") and decides against it (e.g., ``Ah nevermind") the system is able to remove the correct scheduling constraint without additional follow-up questions. 

\begin{figure}[ht]
     \centering
      \includegraphics[width=\textwidth]{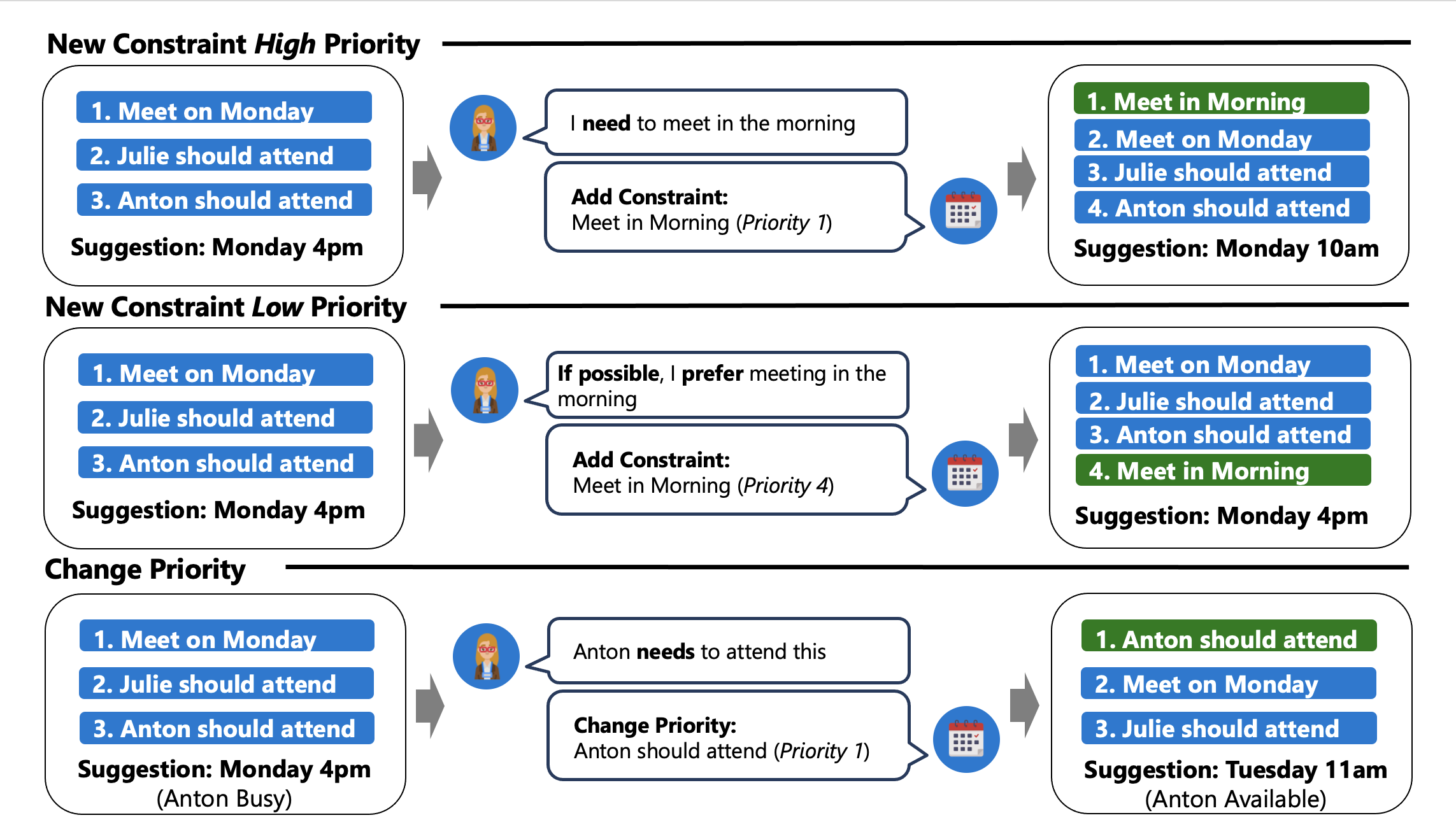}
        \caption{\cl{Three sample actions (represented by rows) taken by the constraint management component to manage constraint priorities. In each example, the left image shows the existing prioritized constraint list and current time suggestion. The middle image shows the new user message and the action taken by the constraint management component. The right image shows the resulting prioritized list of constraints and the new time suggestion. In the first example, a user specifies a new constraint with strong language (i.e., `need') that is translated into a constraint with priority 1. The second example shows the same constraint with weaker language (i.e., `if possible') that is translated into a constraint with priority 4. The final example shows a user specifiying that an unment constraint (Anton's attendance) needs to be met. The constraint manager changes the priority of that constraint to 1. \label{fig:priority_mgmt}}}
        \Description{Three sample actions (represented by rows) taken by the constraint management component to manage constraint priorities. In each example, the left image shows the existing prioritized constraint list and current time suggestion. The middle image shows the new user message and the action taken by the constraint management component. The final image shows the resulting prioritized list of constraints and the new time suggestion. In the first example, a user specifies a new constraint with strong language (i.e., `need') that is translated into a constraint with priority 1. The second example shows the same constraint with weaker language (i.e., `if possible') that is translated into a constraint with priority 4. The final example shows a user specifiying that an unment constraint (Anton's attendance) needs to be met. The constraint manager changes the priority of that constraint to 1.}
\end{figure}

\cl{A key function of the constraint management component is to set the priority of each constraint, and thus maintain the relative ordering of constraints. Upon adding a new constraint, this constraint manager is given access to the current list of scheduling constraints and their priorities and asked to specify the priority of the new constraint. This gives the LLM the flexibility to adjust the priority based on the wording of the new constraint. Based on the chat history, the constraint management component can also decide to change the priority of a constraint (e.g., increasing the probability of satisfying a violated constraint that the user says is important). The system also enables users to change the priority of the constraint during the scheduling process (e.g., if they want a constraint to have higher or lower priority), which helps support preference evolution over time. Figure \ref{fig:priority_mgmt} shows example actions taken by the constraint management component in response to different user messages. A limitation of this priority management functionality is that it remains too opaque to the end user - giving little insight into \textit{why} a constraint was given a certain priority or access to the prioritized list itself. We further discuss the challenges of using LLMs with respect to system transparency in Section \ref{subsec:transparency}.} 

\subsection{Constraint Generation} \label{sec:const_gen}
\begin{figure}[ht]
     \centering
      \includegraphics[width=\textwidth]{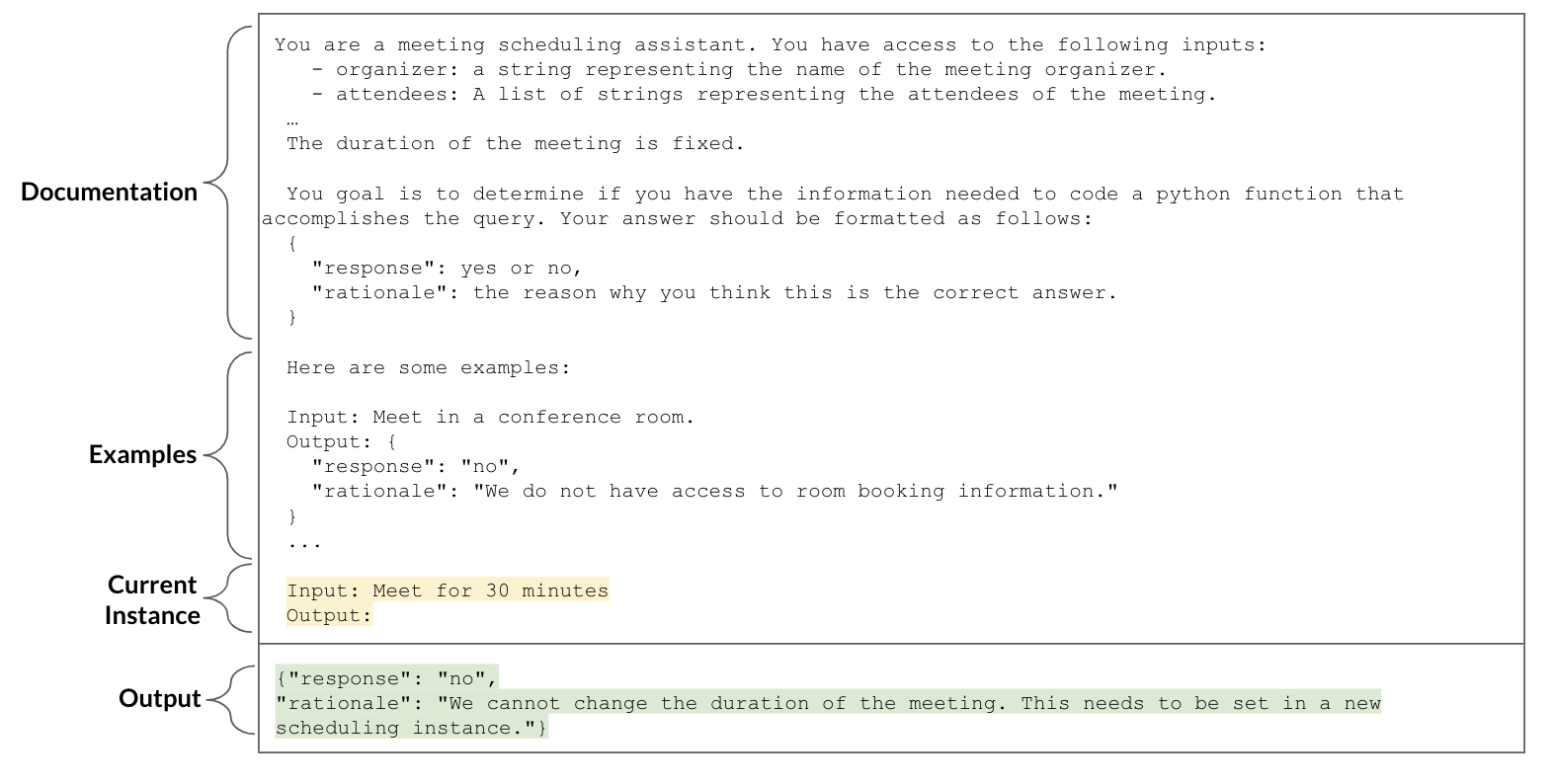}
        \caption{Sample prompt for the information checker component. Orange highlight reflects query-specific inputs. Green highlights the output of the LLM. \label{fig:info_checker_prompt}}       
        \Description{Three input components: Documentation, Examples, and Current Instance as well as an Output component. In this sample, Documentation includes:You are a meeting scheduling assistant. You have access to the following inputs: 
    - organizer: a string representing the name of the meeting organizer. 
    - attendees: A list of strings representing the attendees of the meeting. …The duration of the meeting is fixed. Your goal is to determine if you have the information needed to code a python function that accomplishes the query. Your answer should be formatted as follows:{"response": yes or no,"rationale": the reason why you think this is the correct answer.}}
\end{figure}

\begin{figure}[ht]
     \centering
      \includegraphics[width=\textwidth]{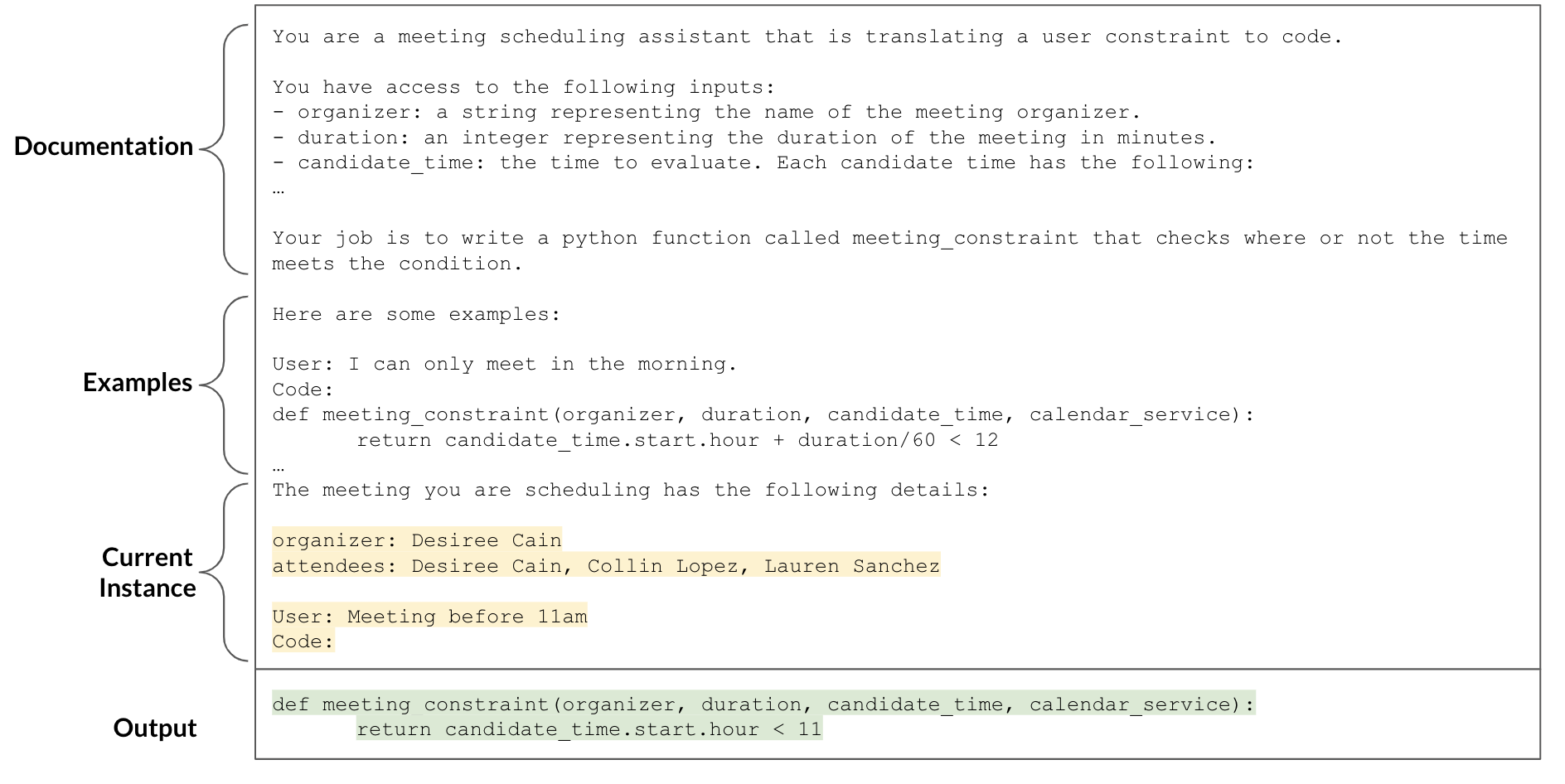}
        \caption{Sample prompt for the constraint generation coder component. Yellow highlight reflects scheduling instance-specific inputs. Green highlights the output of the LLM. \label{fig:coder_prompt}}       
        \Description{Three input components: Documentation, Examples, and Current Instance as well as an Output component. In this sample, Documentation includes: ‘You are a meeting scheduling assistant that is translating a user constraint to code. You have access to the following inputs: -organizer: a string representing the name of the meeting organizer. -duration: an integer representing the duration of the meeting in minutes. -candidate_time: the time to evaluate. Each candidate time has the following: … Your job is to write a python function called meeting_constraint that checks where or not the time meets the condition.’ Examples include: ‘Here are some examples: User: I can only meet in the morning. Code: def meeting_constraint(organizer, duration, candidate_time, calendar_service): return candidate_time.start.hour + duration/60 < 12 …’ Current Instance includes: ‘The meeting you are scheduling has the following details: organizer: Desiree Cain. attendees: Desiree Cain, Collin Lopez, Lauren Sanchez. User: Meeting before 11am. Code:’ Output includes: ‘def meeting_constraint(organizer, duration, candidate_time, calendar_service): return candidate_time.start.hour < 11.’ From Current Instance, the lines organizer: Desiree Cain. attendees: Desiree Cain, Collin Lopez, Lauren Sanchez. User: Meeting before 11am. Code:’ are highlighted in orange. The entire Output is highlighted in green.}
\end{figure}

When the system decides to add a new scheduling constraint, the system calls two LLM-powered components. The first, dubbed \textit{Information Checker}, checks whether the given scheduling constraint can be handled by the system. Given the diversity of user scheduling preferences and the wide range of external information needed to integrate them, it is unlikely that any system will be able to handle all scheduling constraints. The Information Checker acts as a safeguard to check whether the system has sufficient information to be able to handle the constraint. 
\cl{Within the prompt, the information checker is given all information and data sources accessible to the coding module (e.g., attendee information, function to check each user's free/busy time) and asked whether the given constraint (e.g., `I want to meet before I leave for vacation') can be coded with an associated rationale (see Figure \ref{fig:info_checker_prompt} for a sample prompt). If the information checker predicts that the constraint cannot be implemented with the given data sources, the system generates a new observation for the Constraint Manager component that states that the constraint was not added and gives the information checker module's rationale as to why. This enables the constraint manager to query additional information needed to implement the constraint (e.g., ask user to specify when they leave for vacation).} 
If the system can handle the constraint, it is then given to a \textit{Coder} module that translates the scheduling constraint into Python code that can be used to check whether a candidate meeting time meets the constraint. Figure \ref{fig:coder_prompt} shows a sample prompt for generating a new scheduling constraint.  

\subsection{Constraint Programming Solver} \label{sec:cp_solver}

Given a list of scheduling constraints and their associated priorities, the \textit{Meeting Scheduling Problem} aims to find a candidate meeting time with the highest score, defined as the weighted sum of the constraints satisfied. We now briefly formally define the associated constraint programming problem, and show that the problem can be solved efficiently. Let ${\mathcal T}$ be the set of candidate times or a meeting. For instance, for a one-hour meeting on Tuesday, we may list all one-hour blocks starting on the hour or half-hour between 8am and 6pm (i.e., ${\mathcal T} = \{$8-9am, 8:30-9:30am, 9-10am, $\dots$, 5-6pm$\}$). Let $f: {\mathcal T} \rightarrow \{0,1\}$ be a scheduling constraint that maps a candidate time $t \in {\mathcal T}$ to a boolean value representing whether or not the time satisfies it. As outlined in Section \ref{sec:const_gen} these constraints are represented by Python functions as generated by a LLM. As input to the Meeting Scheduling Problem, we have a set of candidate times ${\mathcal T}$, $n$ scheduling constraints ${\mathcal F} = \{f_1, \dots, f_n\}$ each with associated weight (or priority) $w_1, \dots, w_n$. Formally, the goal of the Meeting Scheduling Problem is to solve the optimization problem:
$$
t^* = \text{argmax}_{t \in {\mathcal T}} \sum_{f_i \in {\mathcal F}} w_i f_i(t)
$$

Luckily, for practical meeting scheduling instances, the size of the candidate time set ${\mathcal T}$ is small enough that we can use a brute-force approach to score all candidate times and return the best. For reference, a meeting scheduling instance with 100,000 candidate times and 10,000 scheduling constraints can be solved on a single thread in under 10 seconds. This allows us to avoid the need of a formal constraint programming solver such as Google OR-Tools. Integrating such a solver for more complicated or computationally demanding optimization problems is a promising direction for future work.

In response to feedback from the diary study, the \system~system was also designed to return diverse time suggestions. Instead of returning only a single-time suggestion with the best weighted score, we return a set of $k$ diverse times. Note that simply returning the top $k$ times in terms of score, without considering diversity, can result in similar time suggestions as similar times often have similar properties and scores (e.g., suggesting 9-9:30, 9:30-10, and 10-10:30 if they are the first three available time slots). We define diversity as the sum of the pairwise distance between each time in the returned set $\{t_1, \dots, t_k\}$:
$$
\sum_{i=1}^{k-1} \sum_{j=i+1}^{k} d(t_i, t_j)
$$
where $d(t_1, t_2)$ is a distance function between two suggested times. Motivated by the results of the diary study which highlighted the importance of temporal diversity, we define the distance between two candidate times as the logarithm of their difference in minutes plus one. Finding diverse subsets is known to be a computationally demanding problem \cite{agrawal2009diversifying}. To construct diverse time suggestions we employ a simple greedy algorithm that we found works well in practice. We start by filtering the candidate times to only times within $\epsilon$ of the best score (i.e., to ensure high-quality time suggestions). We select $\epsilon$ to be the smallest value such that there are at least $k$ times that meet the criteria. \cl{In practice, since our scoring system gives integral scores (i.e., because $w_i$ is integer, and $f_i(t)$ is binary), there are often many time slots that receive the same score. Intuitively, in settings with fewer scheduling constraints (e.g., three attendees, must be in the afternoon) many potential meeting times often meet the same or similar subsets of constraints (e.g., times on tuesday or thursday afternoon). Consequently, although our setting of $\epsilon$ only guarantees at least $k$ times are included in the filtered set of candidate times, in most instances there are many more times than $k$ included. This leads to a diverse pool of candidate times that are approximately optimal with respect to the given constraints.} After filtering the times, we greedily construct the set by first selecting the earliest high-scoring time and iteratively adding a new time to the set that maximizes the diversity of the current set. 

To help provide transparency over the suggested meeting times, we return each meeting time with a one-sentence summary of its pros and cons. To generate this summary, we give an LLM the list of scheduling preferences met and unmet by each suggested time and prompt it to generate a one-sentence summary. For instance here is a sample explanation for a time that met two user preferences but failed to include one attendee: \textit{This time is before 11am on Tuesday, but Anton cannot attend}. While LLMs are known to hallucinate and generate text not supported by the inputs~\citep{bubeck2023sparks}, we found that summarizing small lists of preferences \cl{were consistently faithful, and saw no instances of hallucination for this component in the user study.}

\subsection{System Limitations}
This system was designed as a technology probe~\cite{boehner2012probes,hutchinson2003technology} to test the viability of our hybrid LLM and constraint programming framework, not as a full calendar scheduling system. As such, we include only the minimum viable amount of functionality to test the interactive component between the user and the system and exclude any additional functionality that may be present in other calendar tools. When scheduling a meeting, our system requires the meeting characteristics to be input in a form as opposed to a naturalistic chat. \cl{Future research is needed to determine whether these meeting characteristics (or a subset thereof) can be elicited robustly from chat and if that provides a better user experience than this GUI-based approach.} These meeting characteristics are also fixed for the remainder of the interaction (i.e., the user cannot dynamically change the attendees or length of the meeting during chat). The system only supports meetings that begin at increments of 15 minutes (i.e., on the hour, and fifteen-minute increments thereafter). The system is also limited in what it can communicate to users. Specifically, it can communicate suggested meeting times, explanations of those times, and some limited follow-up questions. The system can not interface with external tools or APIs (e.g., sending e-mails to participants, querying weather data, or accessing conference room availability). \cl{We also note that the diversity selection criteria for returning multiple candidate times is fixed and does not allow a user to change what diverse options are meaningful to help their scheduling or specify what constraints they are willing to sacrifice for diverse options (e.g., if a user is willing to compromise on constraints to have meeting time options on different days of the week).}

\section{Quantitative Evaluation of Preference Incorporation}
\label{sec:quant_eval}

To gauge the feasibility of the \system~ hybrid LLM and CP approach for preference incorporation we provide a comprehensive numerical evaluation of two key system components: the information checker, and the coder. 

\subsection{Datasets}
To quantitatively evaluate the system components we create a novel evaluation benchmark that leverages both a synthetic calendar universe and real scheduling preferences captured during the diary study. 
The synthetic calendar universe is used to create sample meeting scheduling scenarios without the use of users' personally identifiable private calendar data. To generate the calendar universe we use GPT4 \cite{openai2023gpt4} to create a synthetic organization with 32 employees spread over 4 distinct teams. We also use the LLM to generate a suite of existing meetings between the employees to populate their calendars. The number of meetings for each employee were commensurate with their position with managers having a higher meeting load, and calibrated to be in line with the self-reported number of attended meetings in the diary study. We also generate a dataset of 75 new meeting instances that represent meetings that need to be scheduled. 

\begin{figure}[!t]
     \centering
      \includegraphics[width=\textwidth]{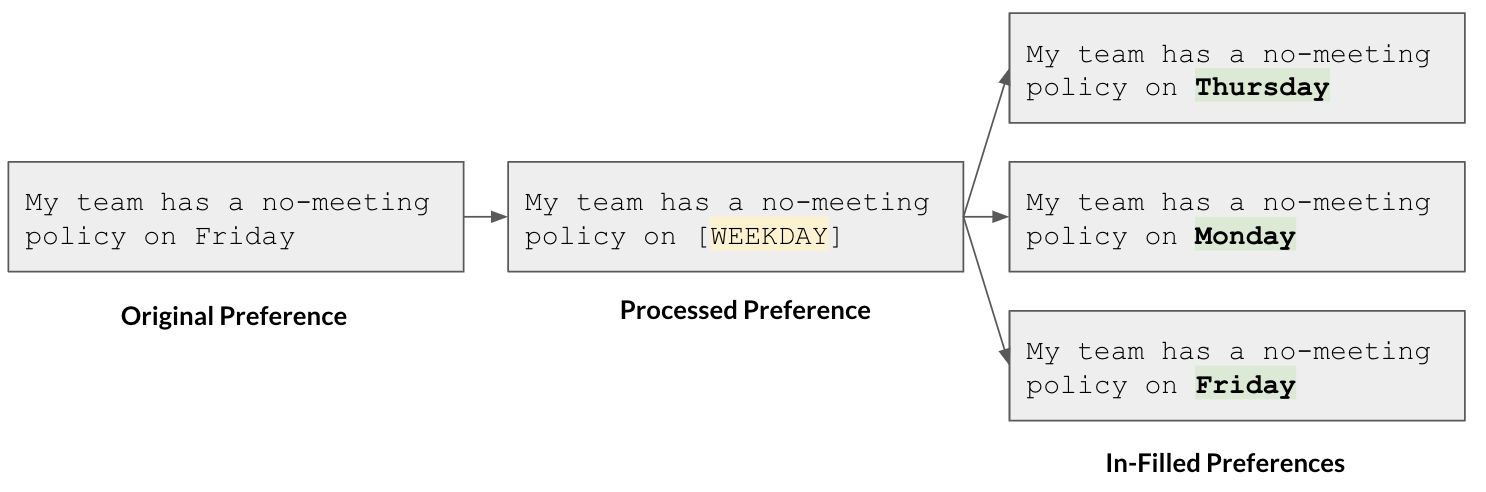}
        \caption{Processing pipeline for preferences from the diary study. \cl{A preference from the diary study (original preference) is pre-processed to replace any specific times, days of the week, or meeting specific information with a special placeholder (e.g., ``[Weekday]'' in diagram). To generate a set of preferences for the dataset, the placeholder in the processed preference is in-filled with either relevant meeting details from the synthetic scheduling instance, or in-filled uniformly at random from a set of allowable settings (i.e., Monday through Friday for the weekday tag).}\label{fig:dataset_process}}      
        \Description{Three components that feed into each other, from Original Preference to Processed Preference to In-Filled Preferences. An example of an Original Preference is ‘My team has a no-meeting policy on Friday.’ This becomes the Processed Preference ‘My team has a no-meeting policy on [WEEKDAY],’ which turns into the In-Filled Preferences ‘My team has a no-meeting policy on [Thursday, Monday, or Friday],’ where Thursday, Monday, and Friday are possible options for [WEEKDAY].}
\end{figure}

To generate a dataset of scheduling preferences we process the results of the diary study. We first extracted all scheduling preferences and constraints elicited during the diary study and removed any inadvertantly shared personally identifiable information. To help generalize the preferences elicited during the diary study we also remove any specific times, days of the week, or meeting specific information (i.e., attendees, project title) and replace them with special placeholders. These placeholders are later in-filled by either sampling the inputs from a uniform distribution over potential inputs (e.g., randomnly sample one of the five different workdays) or entering the relevant meeting characteristic (e.g., inputting the name of the organizer for the meeting instance). For every original preference from the diary study, we generate 3 new preferences by in-filling the placeholder values for new meeting scheduling instances. Figure \ref{fig:dataset_process} shows a sample processing pipeline for a final scheduling preference. 

We categorize all the processed preferences from the diary study into two categories: requests that can be handled given the limitations of the synthetic calendar universe and those that cannot. For instance, since the synthetic universe does not have any facility information, any preference related to the availability of conference rooms is marked as not handled. These categories form the basis of a binary classification dataset, which we call the \textit{Safeguard Dataset}, through which we evaluate the information checker component of the system. We also take the subset of the dataset that can be handled by the current system (i.e., could be implemented in python with available data sources) and use it to form a dataset, called the \textit{Code Generation Dataset}, for the coder component. 

\cl{To summarize the overall dataset generation process we first process the diary study results to create a set of natural language scheduling preferences and use it to create two datasets. The \textit{Safeguard Dataset} is a binary classificaiton dataset that contains natural language constraints and whether or not the current system has sufficient data sources to handle it. The \textit{Code Generation Dataset} contains a set of natural language contraints which can be implemented in python with the available data structures.}

\subsection{Information Checker}
We use the Safeguard dataset to benchmark the performance of current LLMs for the Information Checker component of the \system~system.  We evaluate two different LLMs: GPT-3 (text-davinci-003) \cite{brown2020language} and GPT-4 \cite{openai2023gpt4}. We also evaluate two different phrasing strategies. During initial experiments we noticed that the LLMs struggled to extract correct scheduling preferences when given long sentences with rationale. We add a LLM-powered rephraser component that, given the initial scheduling preference, is asked to rephrase it as a succinct scheduling constraint. We note that this is not necessary in the full implementation of the \system~system as the constraint manager component rephrases the chat message into scheduling constraints. We report the classification accuracy of the models and phrasing strategies on the reference dataset, defined as the fraction of the instances the model correctly predicts can and cannot be handled by the current system.

Table \ref{tab:quant_eval} (Safeguard Accuracy column) summarizes the accuracy for the two different models and prompting strategies. Both models have around 80\% accuracy for the information checking task when using the rephraser. The rephraser does indeed lead to a small increase in performance for both models, highlighting the importance of translating complex user queries into clear requests before generating code. 

\begin{table}
\centering
\caption{Comparison of LLM performance on information checking and code generation components on datasets generated from the results of the diary study. All numbers are reported as percentages.}
\label{tab:quant_eval}
\resizebox{\textwidth}{!}{%
\begin{tabular}{ll|l|lllll}
 &
   &
   &
   &
  \multicolumn{2}{c}{\textbf{Correctness (General)}} &
  \multicolumn{2}{c}{\textbf{Correctness (Example)}} \\ 
     \textbf{LLM} &  \textbf{Rephraser} &   \textbf{Safeguard Accuracy}     &  \textbf{Compilation}      & Precision & Recall & Precision & Recall \\ \toprule
\multirow{2}{*}{GPT-3} & Y & \textbf{81.8\%} & 95.3\% & 95.5\%    & 92.6\% & 94.4\%    & 92.4\% \\
      & N & 77.9\% & 90.7\% & 94.6\%    & 86.0\% & 93.8\%    & 87.2\% \\ \midrule
\multirow{2}{*}{GPT-4} &
  Y &
  79.8\% &
  \textbf{97.2\%} &
  \textbf{95.8\%} &
  \textbf{94.0\%} &
   \textbf{94.7\%} &
   \textbf{93.8\%} \\
      & N & 72.7\% & 93.4\% & 95.2\%    & 89.6\% & 94.2\%    & 90.3\% \\\bottomrule
\end{tabular}
}
\end{table}

\subsection{Code Generation}
To evaluate the coder component of the \system~system we compare the LLM generated code for a single preference to implementations of each function generated by human software developers. To ensure the correctness of the implementations, all reference implementations were checked independently by a different software engineer then the one that implemeneted it. The correct implementations follow a similar processing strategy as the original dataset. An implementation is generated only for each processed preference with placeholders from the diary study, and the placeholders are in-filled for the specific meeting context or sampled values. We evaluate three different metrics for the code generation component:
\begin{itemize}
    \item \textbf{Compilation:} The fraction of functions generated by each component that can successfully be imported by a python interpreter.
    \item \textbf{Correctness - Precision:} Of the code that runs successfully, what fraction of the candidate times marked as satisfying the given preference do in fact meet the preference.
    \item \textbf{Correctness - Recall:} Of the code that runs successfully, what fraction of the candidate times that should be marked as satisfying the given preference are marked as such by the code.
\end{itemize}

To compute the precision and recall of the generated code, we first generate two sets of candidate times for each preference in the code generation dataset. The first set, which we call the general dataset, includes every meeting time of the correct duration over the next 50 days. The second, which we call the example dataset, restricts the candidate times to only times with the correct duration over the time horizon of the synthetic meeting scheduling example which varied from 2 to 14 days depending on the example. We evaluate both sets to account for instances where an error in the code generation component may not appear over a small time horizon. Table \ref{tab:quant_eval} (right side) reports the performance of both models and prompting strategies over the five metrics. Both LLMs are able to generate code that compiles correctly over 95\% of the time and have high precision and recall (over 90\%). This showcases the viability of using LLMs to generate correct python representations of user scheduling preferences.


\section{User Study}
\label{sec:user_study}

To evaluate the feasibility of \system, we conducted a user study with 10 participants who used the \system~user interface to schedule meetings as a member of a fictitious organization. 
In this user study, we evalute the \system~system not as a finished product but as a ``probe'' \new{that inspires design ideas}~\cite{boehner2007hci}. \new{
The concept of using a probe as a provocational or an inspirational instrument to ``open a conversation~\cite{boehner2012probes}'' and to ``find out about
the unknown---to hopefully return with useful or interesting
data~\cite{hutchinson2003technology}'' has been well-established in the HCI literature for many decades~\cite{gaver1999design,graham2008probes,boehner2007hci,hutchinson2003technology}. }
\new{This method invites user participation in the early stages of the design process and to promote flexible and open-ended use with less emphasis on usability evaluation.}
\new{In this user study, we utilize the \system~ as a ``technology probe''---a functional piece of technology artifact presented to the users---with social science, engineering, and design goals as outlined by \cite{hutchinson2003technology}.
Our goal with the technology probe is} to observe the interactive preference elicitation process in real-world scheduling scenarios and to uncover design implications and unforeseen challenges for further improving the system design.
Our study \new{protocol and} tasks are, therefore, intentionally designed to \new{provoke open-ended reactions rather than task successes and to} push the limits of the current system design. 

\subsection{\system~UI}
\begin{figure}[t!]
     \centering
      \includegraphics[width=\textwidth]{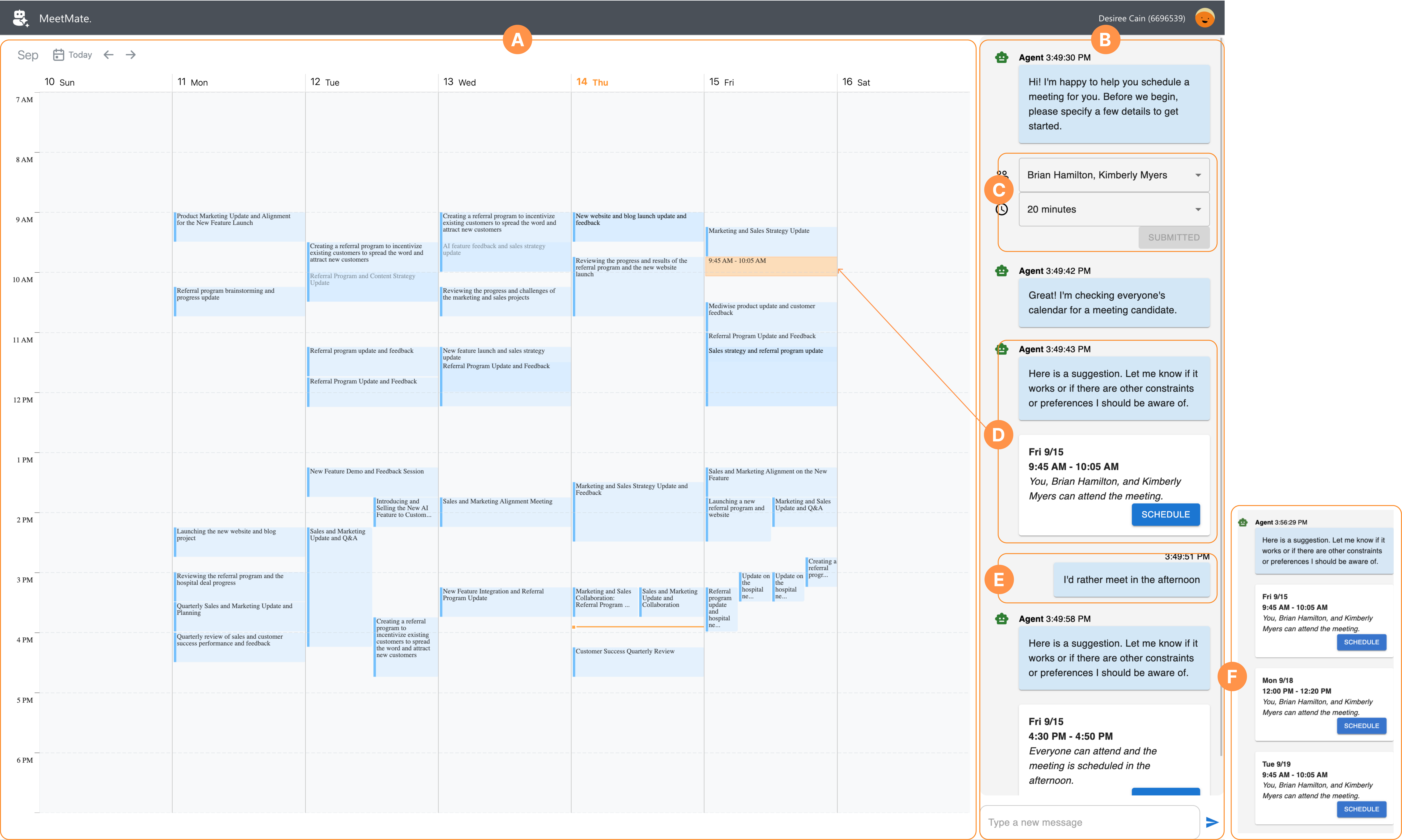}
        \caption{\system~ user interface consists of the calendar component on the left (A) and the chatbot component on the right (B). Users enter meeting scheduling details into a form (C). Upon submission, the system generates an initial time suggestion (D), which users can refine via naturalistic chat (E). The system can also be configured to return multiple diverse time suggestions (F).}
\label{fig:system_ui}
\Description{Two components: (A) Calendar and (B) Chatbot. Chatbot subcomponents can include a (C) Scheduling Form, (D) Initial Time Suggestion, (E) User Input, and (F) Multiple Time Suggestions.}
\end{figure}

We designed a UI for \system~to allow users to interact with the underlying system.
The main goal of the design is to provide a minimal set of capabilities that lets us exercise the underlying system components rather than to provide a full set of functionalities of a typical calendar.
The UI consists of a calendar component on the left (Figure~\ref{fig:system_ui}A) and a chatbot component on the right (Figure~\ref{fig:system_ui}B). The calendar component resembles a typical calendar client and displays the current user's weekly schedule. The chatbot component allows for naturalistic conversations with the system with a chat history and a message box. \cl{All components of \system~(i.e., Constraint Manager, Information Checker, and Coder) used GPT-3 for the user study as it had comparable performance to GPT-4 in the quantitative evaluation outlined in Section \ref{sec:quant_eval} and, at the time of the user study, had a lower latency than using GPT-4.}

The system UI was designed to support only allowable interactions with the underlying components. As such, the user started the interaction by first specifying a list of attendees and the duration of the desired meeting (Figure~\ref{fig:system_ui}C), similar to modern calendaring tools like Microsoft Outlook or Google calendar, and the system responded with an initial suggestion with a short explanation (Figure~\ref{fig:system_ui}E). All meetings were set to be scheduled in a default time horizon of one week. If the suggested time is satisfactory, the user can schedule the meeting by clicking on the ``Schedule'' button. If the suggested time is unsatisfactory, the user can freely interact with the system through the message box (Figure~\ref{fig:system_ui}E). 
Each time a user sends a message, the system responds with a suggested time and/or a message, in which case the user either accepts the suggestion or iterates by interacting with the system. 
The system could be configured to return multiple suggestions at a time (Figure~\ref{fig:system_ui}F). 

\subsection{Study Design}
Each consented participant attended a 1-hour user study session conducted remotely via a video conferencing tool, and the sessions were video-recorded and transcribed for analysis.
During the study, participants were asked to use \system~UI to schedule meetings for a fictitious healthcare software start-up while assuming various personas with different preferences and constraints for meetings. 
In total, we recruited 10 people on a first-come-first-serve basis from a sample that participated in our diary study and agreed to be contacted for future studies. Participants self-reported as working in the roles of a product manager (40\%), manager (20\%), developer (10\%), and researcher (10\%). 60\% of participants reported attending 10-19 meetings per week, with the remaining 40\% attending over 20 meetings.
Each participant was compensated with a \$50 gift card.
The study design and protocol were approved by the leading institution's Institutional Review Board (IRB). 

Each study session consisted of four scheduling tasks crafted to challenge the system's capabilities.
The tasks and the calendars that participants interacted with were set up either to make it difficult for the system to satisfy all preferences or to allow users to engage with the system in unanticipated ways that would potentially break the system.
The first task involved a 40-minute meeting with three other attendees with preferences for it being after 11am, a 30-minute break before the meeting, and allowing for a meeting over lunch, which were preferences that \system~ was designed to support. The task was designed to also test the \textit{Information Checker} by adding a preference to meet on a sunny day, which the system was not designed to support. 
The second task involved scheduling an urgent, next-day, 60-minute meeting with two required attendees and two optional attendees, including the organizer. 
The third task involved scheduling a 60-minute meeting with three other attendees, and the participant was allowed to bring in any meeting preferences or constraints they wanted to exercise to evaluate the system.
The fourth task was identical to the third task, except that the first three tasks returned one suggestion at a time and the fourth task returned three suggestions at a time. \cl{The modality of the meetings (e.g., virtual, in person, hybrid) were not explicitly stated in the user study instructions giving users the maximum freedom to determine what scheduling constraints they wanted to consider.} 
The third task was optional if there was not enough time to complete all four tasks within one hour \cl{(3 out of 10 participants completed it)}. \cl{Note that in all tasks, users were provided with the duration, time frame, and attendees of the meeting they were instructed to schedule.} Each task was considered done when the participant found a satisfactory meeting time or when they expressed that they could not complete the task. With the exception of the first task, all tasks were designed to be challenging to complete with the current system to help expose new challenges within the interactive decision support framework. Participants were not taught about the limitations of the systems and what scenarios were not supported (e.g., calendar question and answering, dynamically changing meeting time). This was to gauge the types of features participants expected in a dynamic setting and to evaluate how well participants could recover from shortcomings in the system.

Throughout the study, participants were asked to verbalize their thoughts (i.e., think-aloud protocol).
During the tasks, the facilitators asked follow-up questions probing the participants' comments or behaviors. 
After each task, we asked participants about their ability to find a suitable meeting time, the ease of expressing preferences, the ease of using the system to schedule a meeting, and their understanding of the limits of the system's abilities. 
We also asked them for overall feedback about their scheduling experience, a comparison between their current scheduling practice and using \system, and any feedback about the number of suggestions provided. 
\new{These questions were used as a starting point of open-ended discussions about the system design and the tasks. It is important to note that, while our study allowed for capturing of quantifiable data such as task success rate, we do not report such metrics because of the inherent open-ended and conversational nature of our study design that make such metrics invalid.}


\subsection{Analysis}
We reviewed the qualitative texts using reflexive thematic analysis~\cite{braun2023thematic}.
The recordings and transcripts were uploaded to Marvin\footnote{https://heymarvin.com/}, and each recorded session was open-coded by three researchers who either facilitated or attended that session on a line-by-line basis, and the codes were iterated and grouped into higher-level themes. 
Throughout and after coding, researchers met to resolve disagreements and iteratively reach a consensus on the overarching themes.

\subsection{Findings}




Throughout our study, we observed the iterative preference elicitation process, between the user and the system, to arrive at a desired scheduling outcome: participants constructed their preferences based on system output (i.e., preference construction), and the system incorporated them into newly generated suggestions for participants to re-evaluate (i.e., preference incorporation).
When our participants verbalized their reasoning, we were able to peek into the mental steps that occurred during these iterations. 

Overall, users found the interactive decision support system easy to use and responsive to their preferences. Specifically, users highlighted the ease at which they could communicate their preferences to the system: ``The natural language input was easy to use. Right. I don't have to learn syntax or much of anything (P6)'', ``I would schedule this way. I'm real partial to chatting, so I'm very used to it (P4)''. Users also remarked how flexible the agent was in responding to their preferences: ``When I picked only the first names it was able to understand and that's a great thing (P9)", ``It totally understood what I meant, so the suggestions it gave me actually works. (P10)'', ``I liked that it was able to meet preferences (P5)". \cl{The system also supported evolving preferences over time. For instance, P2 was able to instruct the system to find a time that could include optional meeting participants at the expense of satisfying a preference for having the meeting the next calendar day.} Multiple users expressed wanting to use the current system further (P1, P4, P8). 

Our findings also underscored the importance of an interactive and contextualized system for decision support. Participants noted that some scheduling preferences were only revealed within the iterative process of scheduling the meeting. For instance, throughout their interview P7 asserted that they prefer scheduling meetings at the first available time for all required attendees. However, at the end of the interview, they revealed that they ``subconsciously prioritize'' to avoid booking out their entire calendar, for instance, ``I don't really want to do [book this time]; they're only available at 5PM. Yeah, I got in at 7 [AM]. I'm not doing a 5:30PM meeting (P7).'' We saw that this fluid and subliminal nature of preferences was a large reason why several participants (P1, P2, P3, P6, P7, P8, P10) preferred receiving multiple suggestions. P6 elaborated on the benefit of having options to help refine preferences: ``Maybe none of them are perfect, but maybe they're close. Is there a compromise?'' Participants also reevaluated preferences based on contextual information in the scheduling tasks. For instance, P10 expressed a preference for a 15 minute break before the meeting if they have been in back-to-back meetings for at least 2 hours. When asked if this is how they typically handle breaks, they described scenarios where they would relax this constraint, including ``if it's just 2 hours and then after that I'm completely free'' or ``if one of those 2 hour meetings is one where I'm just listening in.''

During the preference construction process, we observed two mental steps that helped participants interact with the system -- evaluation and expression.
When presented with scheduling suggestions, we saw that participants \textit{(1) evaluated} the system suggestions to see if their preferences were incorporated and if they matched their desired outcome. In this step, they first tried to interpret the system suggestions, explanations, and messages, by formulating mental models of how the system presents information. Then, they decided whether they were satisfied with the suggestions, or they realized a different constraint and proceeded to construct new preferences. 
Afterward, participants \textit{(2) expressed} their newly constructed preferences. In formulating their preferences as written expressions, they ideated how the system might interpret these preferences. Based on their mental model of how the system operates, participants adjusted their expression of preferences. 
While exercising these mental steps, participants suggested opportunities to make iteration with the system easier.
In this section, we focus our discussion on these two mental steps that support the preference construction process: suggestion evaluation (i.e., how people evaluated the suggestions) and preference expression (i.e., how people expressed their preferences). Subsequently, we present additional considerations for improving the system design.

\subsubsection{Suggestion evaluation}
We observed that understanding the system and its capabilities is necessary to evaluate the quality and appropriateness of the suggestion.
Because the \system~UI was limited in its ability to respond (i.e., only with suggestions, short explanations, and occasional messages), understanding the system capabilities required some trial and error: ``This seems to be with four or five questions I could understand what is it trying to do in the background (P8).''
When participants were unsure about the system, they made incorrect assumptions about the underlying logic or second-guessed themselves. 
For example, when the system generated an inaccurate code for an ``within 24-hour'' constraint by inverting the inequality sign, P4 assumed a different explanation for the system's behavior: ``So it's not really looking at the moment in time of where we are and then backing from the 24 hours to find that window right. Is how it appears to me.'' 
Participants were at times confused about which preferences were satisfied, especially when the system provided the same suggestion: ``I said I must attend and it still gave me the same suggestions. So it didn't remember my initial question or something? (P7)''
Without an adequate understanding of the system, when suggestions did not align with their expectations or were perceived as incorrect, participants gave up their scheduling tasks (P7), did not have trust in the system (P2), and would not use the system (P10). 

How the system presented its explanations affected participants' understanding of system suggestions and which preferences were met. Some participants found that the explanations helped them troubleshoot their preferences: ``It was upfront when it couldn't do something. Right. We asked to take into consideration weather, and it told me right off, I don't have that (P6)''. However, P2 reported that text-based explanations made it difficult to comprehend the suggestion compared to when it is visually positioned in the calendar.
Several participants (P4, P7, P10) simply missed reading the explanations. 
To help with understanding their own preferences, some participants (P6, P7, P8) wanted the system to explicitly specify why some preferences were satisfied or not, and the lack of such adequate explanations was sometimes a reason for participants to engage in trial and error.
For example, P6 mentioned that ``it's not really telling me why it won't put [the meeting] on the 6th'' and proceeded to investigate the reasoning by interacting with the bot. 
Receiving the same suggestion without adequate adjustments to the explanation was especially frustrating to some participants (P3, P5, P6, P7).  
Explaining the suggestion's impact on attendance was a high priority for some participants (P2, P6): ``if I have to schedule with ten people, it's just helpful to understand who cannot attend. (P2)''
Instead of an explanation, P5 suggested the system should ask if the unmet constraint could be relaxed: ``if it couldn't find it on that date, maybe it should have asked, if I'm flexible on the time.'' 
In general, understanding the why behind suggestions helped participants focus their attention on making adjustments to their preferences.

Being presented with multiple suggestions was a popular feature among participants: ``It's in some way more efficient. You're given multiple at once. It saves me a little time right. (P6)''). However, some participants weighed the cost of iterating with the system (e.g., ``if you always have to dive into specifics (P3)'') as a potential barrier to using the system.
Several participants made suggestions for how the system could present its suggestions and explanations in a way that helped them understand the system, reason about the preferences, and arrive at their decisions quickly.
Many participants (P2, P3, P6, P7, P8) suggested a simple explanation when important preferences were not satisfied or different suggestions could not be found (e.g., ``Sorry, this is your only option (P7)''). 
Some participants suggested the system provide diverging or alternate paths as a way to iterate on their preferences quickly. 
For example, P3 suggested making most interactions decision-making oriented, such as juxtaposing different suggestions that satisfy a subset of preferences would help them weigh different options and reduce the number of iterations: ``Because that answers two questions rather than me having to prompt the second question.''
P7 suggested diverging options that prioritized certain preferences over others.
Furthermore, in explaining diverging options, participants suggested explanations that describe candidate meeting times in relation to each other as well as how their preferences were met (e.g., ``if you want it sooner, this is your option (P7).'') or to highlight differences (e.g., ``In which dimensions are they different to make you choose about, okay, this one is better than this one, or this one is closer to what I'm looking for than that one (P1).'').
Participants also suggested visualizing the suggestions in a different way to optimize decision-making. For example, P1 wanted a better visual connection between individual suggestions, and P4 preferred ``swim lanes'' of attendees' availabilities to see the suggestions in context.

In general, participants wanted the system to present suggestions and explanations in a way that would give them more control and agency. 
P2 expressed that the current suggestion-based responses from the system made them feel like ``it's dictating terms and it's not flexible enough for me.''
The \system~system did not support changing the meeting duration after the initial configuration. When the system suggested restarting the scheduling session due to this limitation, P4 explained that ``That will be a hard stop for a lot of people'' because they would not know how to proceed. Instead, P4 suggested helping people ``restate [their] desire'' to regain control. 
Instead of exercising control, P4 found themselves ``accommodating'' to system limitations.


\subsubsection{Preference expression}
Many participants enjoyed the flexibility to express their preferences in natural language: ``I think it definitely allows for more free form. Right. I think it allows a lot more flexibility with being able to just chat with the bot over being locked into a specific clickable response (P6)''. 
Preference expression occurred in two phases: (1) at the start of interacting with the system and (2) in response to a system's suggestion, explanation, or message. 
During the initial setup and communication with the bot, participants suggested other interaction modalities beyond chat-based text. Some participants (P3, P5, P7, P10) thought the ``back-and-forth'' nature of chat could be inefficient, especially when the system's understanding did not align with their own. For example, P5 expressed discontent over needing to rewrite the same preference in four different ways. To circumvent these issues, people suggested the use of a form to capture non-negotiable information upfront, including meeting location (P3, P5, P9), required versus optional attendees (P3, P5, P7, P8, P9), and working hours (P2, P4, P5, P6, P10). Additionally, P1 stated that ``using words to express constraints can get tiring,'' and P7 suggested incorporating audio for blind people. 

When responding to the system's suggestion, participants sometimes struggled to formulate preferences, because they lacked insight into how the system embedded preferences and generated suggestions. 
P1 explicated the importance of establishing a common language with the system and the friction experienced ``because I don't see a lot about what the system understood and how they interpret things.'' Sometimes this confusion was complicated by the limitations of LLMs: P8 acknowledged that longer text can be hard for the system to process---``even when I read it, it takes me a second to really understand what I typed.'' Additionally, participants could not differentiate between when the system did not \textit{understand} a preference and when the system lacked the ability to \textit{act} on a preference, leading participants to spend time ``thinking about how to say what [they] need (P6).'' P10 explained, ``My first reaction was, `I may have asked it wrong. Maybe if I ask it a different way,' which is why I asked it a second time slightly differently.'' Many people (P2, P3, P5, P7, P9, P10) suggested that the system should be able to refine previous suggestions instead of providing entirely new ones. For example, P7 asked the system to ``start the meeting at the bottom of the hour,'' intending to adjust the previous suggestion from 2:45PM to 3:00PM, but the bot returned a new suggestion for 11:00AM. 
However, within a chat log, correctly attributing which preferences are meant to refine a current suggestion versus generate a new suggestion is a challenging problem for both LLMs and humans.

In either phase, we noted that participants assumptions and prior experiences could result in a misalignmnet between the system and user. For example, when P1 requested afternoon options, they had to assert that ``afternoons start at 1PM'' because the system believed afternoons to be ``between 12PM and 3PM.'' 
Participants suggested the system should have knowledge of common scheduling conventions, such as ``the next business day'' (P3, P5, P6) or blocking off lunch (P3, P4). 
Both P2 and P7 mentioned organizational norms such as implicitly understanding that 30-minute meetings ``start 5 minutes after, and we do it for 25 minutes (P2).''
Participants expressed a desire for an intelligent system that could learn from prior behavior and automate what must be manually encoded today (P2, P3, P6, P7, P10), while giving users the ability to adjust learned constraints (P3, P6).

\subsubsection{Other considerations}
Our study uncovered findings that applied beyond the preference elicitation process and could benefit future system design. 
We found that simply supporting preference elicitation and scheduling meetings was not enough and that participants wanted seamless integration of other aspects of meetings to be a part of this process.
For a meeting where the organizer could not attend, P8 suggested support for setting an agenda for others that can attend the meeting: ``since it said I cannot join as part of the schedule itself, it can give me some suggestions about like add agenda.''
P3, who self-described as having many meetings, wanted the system to support setting agendas, scheduling breakouts, pulling action items from prior meetings, or pulling in relevant documents.
Completing the meeting scheduling experience also included searching for available office or meeting space (P3, P6), incorporating commute or transition time (P8, P10), and accommodating multiple time zones and business hours (P8).
We also found that, since attendees are a core part of scheduling meetings, participants (P2, P7, P8, P10) sought a feature that allows searching and resolving attendees as part of the UI.



Leveraging LLMs for incorporating user preferences presented both benefits and challenges. 
As we already described, LLMs enabled all of the participants to freely interact with the system to express their preferences and questions. 
The system was especially good at handling misspellings that enhanced user experience: ``I'm satisfied to verify my hypothesis that it's kind of resilient to me being sloppy while typing. So that's a neat thought about experience. (P1)''
\cl{However, the system also hallucinated or made errors in the LLM components of the system. Table \ref{tab:hallucination} summarizes the hallucinations encountered during the user study, grouped by the type of error made. Overall we found 14 instances of hallucinations during the user study. The majority of errors occured in the Coder component of the system, with the LLM either implementing constraints incorrectly (e.g., by inverting the inequality sign) or by using a different definition of a term than the user expected (e.g., a different understanding of when the afternoon starts). While we expect the former kind of error can be mitigated with improved prompting or with recent advances in the ability of code generating LLMs, the latter errors emphasizes a need for the system to proactively resolve ambiguity with the user. The Information Checker and Constraint Manager also made a small number of errors while deciding whether a given constraint could be coded, or when selecting an action to take respectively. In all of these error cases, the participants were not given enough information about the underlying system (e.g., generated code, list of constraints) to troubleshoot. 
}


\begin{table}
\centering
\caption{\cl{Summary of LLM hallucinations during user study. \# refers to the number of instances where that type of error occured.}} 
\label{tab:hallucination}
\begin{tabular}{p{1.5cm}|p{4cm}|p{0.5cm}|p{7cm}}
     Module & Error Type & \# & Example \\ \toprule
     \multirow{ 2}{*}{Coder} & \textbf{Logic Error.} 
     
     \textit{Python Function implements incorrect code.}& 6& 
     \textbf{Input:} The meeting needs to start after 11am
     
     \textbf{Output:} Code flips inequality sign when checking if meeting occurs \textit{after} 11am:
     \begin{verbatim}
def meeting_constraint(candidate_time):
    return candidate_time.start.hour < 11
\end{verbatim}
\\
     & \textbf{Definition Error.} 
     
     \textit{Misalignment between LLM and user def. of term.}& 2& 
     \textbf{Input:} I want to meet in the afternoon
     
     \textbf{Output:} LLM interprets ``meeting is in the afternoon'' as the meeting ends after 12pm (user expects it to \textit{start} after 1pm):
     \begin{verbatim}
def meeting_constraint(candidate_time):    
    return candidate_time.end.hour >= 12
     \end{verbatim}\vspace{-0.5cm}\\ \midrule
    Information Checker & \textbf{Logic Error.} 
    
    \textit{Incorrect classification of given constraint.} & 3 
    & \textbf{Input: } I want to meet during 2-3pm this week because its sunny in Seattle
    
    \textbf{Output: } LLM states that given constraint  \textit{cannot be implemented} because the LLM ``does not have access to weather data''\vspace{0.3cm} \\ \midrule
     \multirow{ 2}{2cm}{Constraint Manager} & \textbf{Logic Error.} 
     
     \textit{Incorrect action taken given user input.}& 1& 
     \textbf{Input: } Move the meeting ten minutes after the hour.

     \textbf{Output: } Constraint Manager changes the priority that Desiree Cain can attend the meeting to one. \vspace{0.3cm}
     \\
     & \textbf{Multi-Action Error.} 
     
     \textit{Manager only responds to subset of request.}& 2& 
    \textbf{Input: } I need something the next calendar day. Jennifer and myself are optional for the meeting.

    \textbf{Output: } Constraint Manager lowers the priority that Jennifer and Dana Nguyen attend the meeting, but ignores next calendar day request.
\\
\bottomrule
\end{tabular}
\end{table}


\section{Discussion}
Eliciting preferences in the context of decision support is a challenging task that requires systems that can help users construct their preferences in an interactive contextual manner. In this paper we introduced a novel framework for interactive decision support that leverages LLMs and constraint programming to provide a flexible naturalistic interface. We studied this framework in the context of meeting scheduling, a complex spatial-temporal decision making process. To better understand users' preference construction process during meeting scheduling, we ran a diary study to capture contextualized scheduling preferences and characterize preference construction in the context of meeting scheduling. The results revealed the diversity of contextual user scheduling preferences and the need for a flexible interactive system that can accommodate them.

We used these preferences to design an interactive system for preference incorporation that can refine suggested meeting times in response to user feedback. The system capitalized on the flexibiliy of LLMs to embed natural language scheduling preferences into Python functions that could then be leveraged by a CP solver to generate refined time suggestions. To validate the technical feasibility of this system for preference incorporation we evaluated it on a quantitative benchmark dataset comprised of results from the diary study. The results showed the viability of this hybrid framework, with the LLM components being able to generate valid python code with high precision and recall with respect to the correct underlying scheduling preference. To evaluate the broader interaction between user's internal preference construction process and the preference incorporation system we ran a small scale user study. While users found the system intuitive and lauded its flexibility, the study highlighted key challenges in the design of interactive decision support systems. Namely, users found it difficult to use the system without an understanding of the underlying system and the ability to correct system mistakes. Within the context of meeting scheduling, users also noted that communication by chat could at times be slow and burdensome and recommended features that could allow them to get the feedback and preferences of additional attendees. The remainder of this section elaborates on these important design implications for future interactive decision support systems and discusses limitations of our work.

\subsection{Implications for Design}
Throughout the user study, participants highlighted expectations, pain points, and suggestions regarding the \system~system. We consolidate these suggestions to provide design recommendations and considerations for future interactive decision support systems. 

\subsubsection{System transparency} \label{subsec:transparency}

The \system~system offers concise summaries of the pros and cons of suggested meeting times (e.g., ``This time is before 11am on Tuesday, but Anton cannot attend'').
Through our user study, we discovered that additional transparency measures could enhance user comprehension of the system's operations and, ultimately, improve decision-making in finding suitable meeting times.
This aligns with the concept of \textit{explainable AI}~\citep{arrieta2020explainable,adadi2018peeking}, which highlights the importance of transparency interventions in enhancing system acceptance, trust, human control, and efficiency in working with the system~\citep{langer2021we}---topics that were addressed during our system evaluation. According to \citet{lazar2022legitimacy}, the overarching goal of system transparency is ``justified understanding,'' tailored to the unique goals and capabilities of stakeholders.
This underscores the need to effectively convey relevant information that users seek and present it in a comprehensible manner.
Our user study suggested that incorporating explanations about the system's internal workings, including its assumptions and accessible information, could help users grasp how the system functions and use it more effectively.
Such explanations may also assist users in understanding why the system made specific suggestions, which is particularly vital when the suggestions deviate from user expectations~\citep{adadi2018peeking}.

Building on this, \citet{lebovitz2022engage} have emphasized the significance of interrogation practices that allow users to connect their own knowledge with AI predictions.
Other research has explored interventions that aid humans in reasoning about the information available to AI systems~\citep{holstein2023toward}.
These interventions may be valuable in our system, enabling users to learn more about what is feasible and what is not as they interact with the system and rely on its recommendations~\citep{lee2004trust,schoeffer2023interdependence}. Finally, when the \system~system presents multiple time suggestions, incorporating \textit{contrastive explanations}~\citep{miller2019explanation}---explanations that compare different time options---could be a valuable addition.
This approach can help users make more informed choices by highlighting the differences, including pros and cons, between various suggestions.

\subsubsection{Human-in-the-loop error recovery}
Multiple participants in the user study noted inconsistencies between their internal preferences and how the system embedded them (e.g., different time ranges for afternoon), however the current system made it challenging for users to correct implicit assumptions made by the system. This was also validated numerically in the quantitative evaluation where even the current state-of-the-art LLM, GPT-4, was unable to have perfect precision, recall, or safeguard accuracy. Given the potential for errors, interactive decision support systems need intuitive ways for users to detect, diagnose, and correct mistakes from the system~\citep{schemmer2023appropriate,schoeffer2024explanations}. This presents a challenging technical problem within the current framework as understanding the underlying representation of user preferences, Python functions, presents a high technical barrier for most users. Future work is needed to bridge this technical gap and provide easy accessible ways to diagnose and correct model error. \cl{Relatedly, some higher-skilled users may want deeper access (e.g., the ability to directly edit constraint functions, akin to scripting in Excel) than the current system provides. Determining how to cater to these different levels of expertise is an important open question for future research.} 

\subsubsection{Reducing elicitation time} 
While the system enabled users to express actionable preferences directly to the system, many users noted the back-and-forth effort of chat to be inefficient and burdensome. Users also voiced frustrations with the system being unaware of scheduling conventions such as lunch times. To help reduce the amount of negotiation time between users and the system to find a meeting time, future systems need to be able to incorporate consistent user preferences and scheduling conventions so that users can focus on fine-tuning suggestions based on new contextual preferences. For instance, this could involve additional form options to capture non-negotiable meeting characteristics like working hours and lunch times as suggested by a number of participants. This could also involve leveraging information about past interactions with the system, including retaining preferences from previous scheduling instances to warm-start the current list of preferences. However, future work is needed to understand how to best leverage preferences across sessions as previous work have shown (scheduling) preferences to be highly transient and contextual \cite{simon2008transience, baharcscw}.

\subsubsection{Multi-user preference construction}
Both the diary study and the user study highlighted the limitations of approaching meeting scheduling solely from the organizer's perspective. Participants noted wanting to query attendees for information such as scheduling preferences, timezone, calendar availability, and feedback on the suggested meeting time. Future work is needed to extend this framework for interactive decision support to a distributed setting where meetings can be scheduled collaboratively between an AI system and multiple attendees. However, such a distributed system presents a number of technical challenges in terms of how to aggregate and weight preferences across users, and how to deal with asynchronous communication between multiple attendees. While previous work has studied this problem in static settings using LLMs \cite{bakker2022fine}, future work is needed to adapt these methods to iterative settings such as meeting scheduling. \cl{Another concern in multi-user settings is the management of information leakage between users. For instance, a primary concern is ensuring undisclosed preferences, sensitive features, and private calendar details are not inadvertently shared when negotiating meeting times between multiple users. While there has been some work exploring privacy concerns in leveraging LLMs \cite{brown2022does}, future work is needed to analyze the trade-offs between privacy and efficacy of multi-user meeting scheduling and negotiation.}

\subsection{Limitations}
While our diary study included participants with a diverse set of roles, they were all employees of the same large technology company. Future work is needed to see if our findings on contextual scheduling preferences generalize beyond the technology sector. The study was also conducted during the summer, which may have limited our ability to capture seasonal variations in scheduling preferences (e.g., those related to the school year or weather). Our user study utilized synthetic scheduling tasks on fake calendar data. While one task required users to use their own scheduling preferences, the manufactured nature of the task might have elicited different preferences than a real scheduling task on the user's own calendar data. \new{Our decision to explore the \system~ as a technology probe limits our ability to draw quantitative conclusions about the usability or effectiveness of the system.} We also highlight the limitations of the scale of the user study. Future work is needed to do larger scale analysis of the system's usability with personal contextual scheduling tasks. We investigated our new framework through the lens of meeting scheduling \cl{ for information workers}, an important but narrow representative task for decision support. \cl{We note that other scheduling scenarios (e.g., coordinating calendars in a social group or academic setting) present a different set of implicit assumptions, scheduling preferences, and potential users that may limit the applicability of our findings.} Future work is needed to evaluate the framework in a broader class of decision making settings. Furthermore, our study focused on the perspective of the meeting organizer without an active involvement of meeting attendees and consideration of their preferences, and we encourage others to build on our framework and the system to explore distributed settings involving multiple stakeholders.

\section{Conclusion}


In this paper, we investigated a new hybrid LLM and optimization framework for interactive decision support. We applied this framework to meeting scheduling, highlighting a number of important design implications for future systems. Our framework could be used in 
other interactive decision support applications such as personalized desk booking, route planning, and supply chain management. Incorporating more sophisticated optimization tools such as commercial solvers (e.g., Gurobi, CPLEX) is a promising future direction for more complicated decision making settings. Interactive decision support is an inherently cross-disciplinary challenge that requires expertise from optimization, human-computer interaction, and machine learning. We encourage others to continue exploring this challenging new approach for decision support.

\bibliographystyle{ACM-Reference-Format}
\bibliography{main}


\begin{thebibliography}{103}


\ifx \showCODEN    \undefined \def \showCODEN     #1{\unskip}     \fi
\ifx \showDOI      \undefined \def \showDOI       #1{#1}\fi
\ifx \showISBNx    \undefined \def \showISBNx     #1{\unskip}     \fi
\ifx \showISBNxiii \undefined \def \showISBNxiii  #1{\unskip}     \fi
\ifx \showISSN     \undefined \def \showISSN      #1{\unskip}     \fi
\ifx \showLCCN     \undefined \def \showLCCN      #1{\unskip}     \fi
\ifx \shownote     \undefined \def \shownote      #1{#1}          \fi
\ifx \showarticletitle \undefined \def \showarticletitle #1{#1}   \fi
\ifx \showURL      \undefined \def \showURL       {\relax}        \fi
\providecommand\bibfield[2]{#2}
\providecommand\bibinfo[2]{#2}
\providecommand\natexlab[1]{#1}
\providecommand\showeprint[2][]{arXiv:#2}

\bibitem[Abdin et~al\mbox{.}(2024)]%
        {abdin2023kitab}
\bibfield{author}{\bibinfo{person}{Marah~I Abdin}, \bibinfo{person}{Suriya
  Gunasekar}, \bibinfo{person}{Varun Chandrasekaran}, \bibinfo{person}{Jerry
  Li}, \bibinfo{person}{Mert Yuksekgonul}, \bibinfo{person}{Rahee~Ghosh
  Peshawaria}, \bibinfo{person}{Ranjita Naik}, {and} \bibinfo{person}{Besmira
  Nushi}.} \bibinfo{year}{2024}\natexlab{}.
\newblock \showarticletitle{KITAB: Evaluating LLMs on constraint satisfaction
  for information retrieval}. In \bibinfo{booktitle}{\emph{The Twelfth
  International Conference on Learning Representations}}.
\newblock


\bibitem[Achiam et~al\mbox{.}(2023)]%
        {openai2023gpt4}
\bibfield{author}{\bibinfo{person}{Josh Achiam}, \bibinfo{person}{Steven
  Adler}, \bibinfo{person}{Sandhini Agarwal}, \bibinfo{person}{Lama Ahmad},
  \bibinfo{person}{Ilge Akkaya}, \bibinfo{person}{Florencia~Leoni Aleman},
  \bibinfo{person}{Diogo Almeida}, \bibinfo{person}{Janko Altenschmidt},
  \bibinfo{person}{Sam Altman}, \bibinfo{person}{Shyamal Anadkat},
  {et~al\mbox{.}}} \bibinfo{year}{2023}\natexlab{}.
\newblock \showarticletitle{GPT-4 technical report}.
\newblock \bibinfo{journal}{\emph{arXiv preprint arXiv:2303.08774}}
  (\bibinfo{year}{2023}).
\newblock


\bibitem[Adadi and Berrada(2018)]%
        {adadi2018peeking}
\bibfield{author}{\bibinfo{person}{Amina Adadi} {and} \bibinfo{person}{Mohammed
  Berrada}.} \bibinfo{year}{2018}\natexlab{}.
\newblock \showarticletitle{Peeking inside the black-box: A survey on
  explainable artificial intelligence (XAI)}.
\newblock \bibinfo{journal}{\emph{IEEE Access}}  \bibinfo{volume}{6}
  (\bibinfo{year}{2018}), \bibinfo{pages}{52138--52160}.
\newblock


\bibitem[Agrawal et~al\mbox{.}(2009)]%
        {agrawal2009diversifying}
\bibfield{author}{\bibinfo{person}{Rakesh Agrawal}, \bibinfo{person}{Sreenivas
  Gollapudi}, \bibinfo{person}{Alan Halverson}, {and} \bibinfo{person}{Samuel
  Ieong}.} \bibinfo{year}{2009}\natexlab{}.
\newblock \showarticletitle{Diversifying search results}. In
  \bibinfo{booktitle}{\emph{Proceedings of the Second ACM International
  Conference on Web Search and Data Mining}}. \bibinfo{pages}{5--14}.
\newblock


\bibitem[Aloysius et~al\mbox{.}(2006)]%
        {aloysius2006user}
\bibfield{author}{\bibinfo{person}{John~A Aloysius}, \bibinfo{person}{Fred~D
  Davis}, \bibinfo{person}{Darryl~D Wilson}, \bibinfo{person}{A~Ross Taylor},
  {and} \bibinfo{person}{Jeffrey~E Kottemann}.}
  \bibinfo{year}{2006}\natexlab{}.
\newblock \showarticletitle{User acceptance of multi-criteria decision support
  systems: The impact of preference elicitation techniques}.
\newblock \bibinfo{journal}{\emph{European Journal of Operational Research}}
  \bibinfo{volume}{169}, \bibinfo{number}{1} (\bibinfo{year}{2006}),
  \bibinfo{pages}{273--285}.
\newblock


\bibitem[Ardissono et~al\mbox{.}(2003)]%
        {ardissono2003framework}
\bibfield{author}{\bibinfo{person}{Liliana Ardissono},
  \bibinfo{person}{Alexander Felfernig}, \bibinfo{person}{Gerhard Friedrich},
  \bibinfo{person}{Anna Goy}, \bibinfo{person}{Dietmar Jannach},
  \bibinfo{person}{Giovanna Petrone}, \bibinfo{person}{Ralph Schafer}, {and}
  \bibinfo{person}{Markus Zanker}.} \bibinfo{year}{2003}\natexlab{}.
\newblock \showarticletitle{A framework for the development of personalized,
  distributed web-based configuration systems}.
\newblock \bibinfo{journal}{\emph{AI Magazine}} \bibinfo{volume}{24},
  \bibinfo{number}{3} (\bibinfo{year}{2003}), \bibinfo{pages}{93--93}.
\newblock


\bibitem[Arrieta et~al\mbox{.}(2020)]%
        {arrieta2020explainable}
\bibfield{author}{\bibinfo{person}{Alejandro~Barredo Arrieta},
  \bibinfo{person}{Natalia D{\'\i}az-Rodr{\'\i}guez}, \bibinfo{person}{Javier
  Del~Ser}, \bibinfo{person}{Adrien Bennetot}, \bibinfo{person}{Siham Tabik},
  \bibinfo{person}{Alberto Barbado}, \bibinfo{person}{Salvador Garc{\'\i}a},
  \bibinfo{person}{Sergio Gil-L{\'o}pez}, \bibinfo{person}{Daniel Molina},
  \bibinfo{person}{Richard Benjamins}, {et~al\mbox{.}}}
  \bibinfo{year}{2020}\natexlab{}.
\newblock \showarticletitle{Explainable artificial intelligence (XAI):
  Concepts, taxonomies, opportunities and challenges toward responsible AI}.
\newblock \bibinfo{journal}{\emph{Information Fusion}}  \bibinfo{volume}{58}
  (\bibinfo{year}{2020}), \bibinfo{pages}{82--115}.
\newblock


\bibitem[Bakker et~al\mbox{.}(2022)]%
        {bakker2022fine}
\bibfield{author}{\bibinfo{person}{Michiel Bakker}, \bibinfo{person}{Martin
  Chadwick}, \bibinfo{person}{Hannah Sheahan}, \bibinfo{person}{Michael
  Tessler}, \bibinfo{person}{Lucy Campbell-Gillingham}, \bibinfo{person}{Jan
  Balaguer}, \bibinfo{person}{Nat McAleese}, \bibinfo{person}{Amelia Glaese},
  \bibinfo{person}{John Aslanides}, \bibinfo{person}{Matt Botvinick},
  {et~al\mbox{.}}} \bibinfo{year}{2022}\natexlab{}.
\newblock \showarticletitle{Fine-tuning language models to find agreement among
  humans with diverse preferences}.
\newblock \bibinfo{journal}{\emph{Advances in Neural Information Processing
  Systems}}  \bibinfo{volume}{35} (\bibinfo{year}{2022}),
  \bibinfo{pages}{38176--38189}.
\newblock


\bibitem[Berry et~al\mbox{.}(2007)]%
        {berry2007balancing}
\bibfield{author}{\bibinfo{person}{Pauline Berry}, \bibinfo{person}{Melinda
  Gervasio}, \bibinfo{person}{Bart Peintner}, {and} \bibinfo{person}{Neil
  Yorke-Smith}.} \bibinfo{year}{2007}\natexlab{}.
\newblock \bibinfo{booktitle}{\emph{Balancing the needs of personalization and
  reasoning in a user-centric scheduling assistant}}.
\newblock \bibinfo{type}{{T}echnical {R}eport}.
  \bibinfo{institution}{Artificial Intelligence Center, SRI International}.
\newblock


\bibitem[Berry et~al\mbox{.}(2011)]%
        {berry2011ptime}
\bibfield{author}{\bibinfo{person}{Pauline~M Berry}, \bibinfo{person}{Melinda
  Gervasio}, \bibinfo{person}{Bart Peintner}, {and} \bibinfo{person}{Neil
  Yorke-Smith}.} \bibinfo{year}{2011}\natexlab{}.
\newblock \showarticletitle{PTIME: Personalized assistance for calendaring}.
\newblock \bibinfo{journal}{\emph{ACM Transactions on Intelligent Systems and
  Technology (TIST)}} \bibinfo{volume}{2}, \bibinfo{number}{4}
  (\bibinfo{year}{2011}), \bibinfo{pages}{1--22}.
\newblock


\bibitem[Bertsimas and Tsitsiklis(1997)]%
        {bertsimas1997introduction}
\bibfield{author}{\bibinfo{person}{Dimitris Bertsimas} {and}
  \bibinfo{person}{John~N Tsitsiklis}.} \bibinfo{year}{1997}\natexlab{}.
\newblock \bibinfo{booktitle}{\emph{Introduction to linear optimization}}.
  Vol.~\bibinfo{volume}{6}.
\newblock \bibinfo{publisher}{Athena Scientific}, \bibinfo{address}{Belmont,
  MA}.
\newblock


\bibitem[Bettman et~al\mbox{.}(1998)]%
        {bettman1998constructive}
\bibfield{author}{\bibinfo{person}{James~R Bettman},
  \bibinfo{person}{Mary~Frances Luce}, {and} \bibinfo{person}{John~W Payne}.}
  \bibinfo{year}{1998}\natexlab{}.
\newblock \showarticletitle{Constructive consumer choice processes}.
\newblock \bibinfo{journal}{\emph{Journal of Consumer Research}}
  \bibinfo{volume}{25}, \bibinfo{number}{3} (\bibinfo{year}{1998}),
  \bibinfo{pages}{187--217}.
\newblock


\bibitem[Boehner et~al\mbox{.}(2012)]%
        {boehner2012probes}
\bibfield{author}{\bibinfo{person}{Kirsten Boehner}, \bibinfo{person}{William
  Gaver}, {and} \bibinfo{person}{Andy Boucher}.}
  \bibinfo{year}{2012}\natexlab{}.
\newblock \showarticletitle{14 probes}.
\newblock \bibinfo{journal}{\emph{Inventive Methods}}  \bibinfo{volume}{185}
  (\bibinfo{year}{2012}), \bibinfo{pages}{185--201}.
\newblock


\bibitem[Boehner et~al\mbox{.}(2007)]%
        {boehner2007hci}
\bibfield{author}{\bibinfo{person}{Kirsten Boehner}, \bibinfo{person}{Janet
  Vertesi}, \bibinfo{person}{Phoebe Sengers}, {and} \bibinfo{person}{Paul
  Dourish}.} \bibinfo{year}{2007}\natexlab{}.
\newblock \showarticletitle{How HCI interprets the probes}. In
  \bibinfo{booktitle}{\emph{Proceedings of the SIGCHI Conference on Human
  Factors in Computing Systems}}. \bibinfo{pages}{1077--1086}.
\newblock


\bibitem[Braun et~al\mbox{.}(2023)]%
        {braun2023thematic}
\bibfield{author}{\bibinfo{person}{Virginia Braun}, \bibinfo{person}{Victoria
  Clarke}, {and} \bibinfo{person}{Nikki Hayfield}.}
  \bibinfo{year}{2023}\natexlab{}.
\newblock \showarticletitle{Thematic analysis: A reflexive approach}.
\newblock In \bibinfo{booktitle}{\emph{Qualitative Psychology: A Practical
  Guide to Research Methods}}. \bibinfo{publisher}{SAGE Publications},
  \bibinfo{address}{New York, NY}.
\newblock


\bibitem[Brown et~al\mbox{.}(2022)]%
        {brown2022does}
\bibfield{author}{\bibinfo{person}{Hannah Brown}, \bibinfo{person}{Katherine
  Lee}, \bibinfo{person}{Fatemehsadat Mireshghallah}, \bibinfo{person}{Reza
  Shokri}, {and} \bibinfo{person}{Florian Tram{\`e}r}.}
  \bibinfo{year}{2022}\natexlab{}.
\newblock \showarticletitle{What does it mean for a language model to preserve
  privacy?}. In \bibinfo{booktitle}{\emph{Proceedings of the 2022 ACM
  Conference on Fairness, Accountability, and Transparency}}.
  \bibinfo{pages}{2280--2292}.
\newblock


\bibitem[Brown et~al\mbox{.}(2020)]%
        {brown2020language}
\bibfield{author}{\bibinfo{person}{Tom Brown}, \bibinfo{person}{Benjamin Mann},
  \bibinfo{person}{Nick Ryder}, \bibinfo{person}{Melanie Subbiah},
  \bibinfo{person}{Jared~D Kaplan}, \bibinfo{person}{Prafulla Dhariwal},
  \bibinfo{person}{Arvind Neelakantan}, \bibinfo{person}{Pranav Shyam},
  \bibinfo{person}{Girish Sastry}, \bibinfo{person}{Amanda Askell},
  {et~al\mbox{.}}} \bibinfo{year}{2020}\natexlab{}.
\newblock \showarticletitle{Language models are few-shot learners}.
\newblock \bibinfo{journal}{\emph{Advances in Neural Information Processing
  Systems}}  \bibinfo{volume}{33} (\bibinfo{year}{2020}),
  \bibinfo{pages}{1877--1901}.
\newblock


\bibitem[Brzozowski et~al\mbox{.}(2006)]%
        {brzozowski2006grouptime}
\bibfield{author}{\bibinfo{person}{Mike Brzozowski}, \bibinfo{person}{Kendra
  Carattini}, \bibinfo{person}{Scott~R Klemmer}, \bibinfo{person}{Patrick
  Mihelich}, \bibinfo{person}{Jiang Hu}, {and} \bibinfo{person}{Andrew~Y Ng}.}
  \bibinfo{year}{2006}\natexlab{}.
\newblock \showarticletitle{groupTime: Preference based group scheduling}. In
  \bibinfo{booktitle}{\emph{Proceedings of the SIGCHI Conference on Human
  Factors in Computing Systems}}. \bibinfo{pages}{1047--1056}.
\newblock


\bibitem[Bubeck et~al\mbox{.}(2023)]%
        {bubeck2023sparks}
\bibfield{author}{\bibinfo{person}{S{\'e}bastien Bubeck},
  \bibinfo{person}{Varun Chandrasekaran}, \bibinfo{person}{Ronen Eldan},
  \bibinfo{person}{Johannes Gehrke}, \bibinfo{person}{Eric Horvitz},
  \bibinfo{person}{Ece Kamar}, \bibinfo{person}{Peter Lee},
  \bibinfo{person}{Yin~Tat Lee}, \bibinfo{person}{Yuanzhi Li},
  \bibinfo{person}{Scott Lundberg}, {et~al\mbox{.}}}
  \bibinfo{year}{2023}\natexlab{}.
\newblock \showarticletitle{Sparks of artificial general intelligence: Early
  experiments with GPT-4}.
\newblock \bibinfo{journal}{\emph{arXiv preprint arXiv:2303.12712}}
  (\bibinfo{year}{2023}).
\newblock


\bibitem[Carenini and Poole(2002)]%
        {carenini2002constructed}
\bibfield{author}{\bibinfo{person}{Giuseppe Carenini} {and}
  \bibinfo{person}{David Poole}.} \bibinfo{year}{2002}\natexlab{}.
\newblock \showarticletitle{Constructed preferences and value-focused thinking:
  Implications for AI research on preference elicitation}. In
  \bibinfo{booktitle}{\emph{AAAI Workshop on Preferences in AI and CP: Symbolic
  Approaches}}. \bibinfo{pages}{1--10}.
\newblock


\bibitem[Carenini et~al\mbox{.}(2003)]%
        {carenini2003towards}
\bibfield{author}{\bibinfo{person}{Giuseppe Carenini},
  \bibinfo{person}{Jocelyin Smith}, {and} \bibinfo{person}{David Poole}.}
  \bibinfo{year}{2003}\natexlab{}.
\newblock \showarticletitle{Towards more conversational and collaborative
  recommender systems}. In \bibinfo{booktitle}{\emph{Proceedings of the 8th
  International Conference on Intelligent User Interfaces}}.
  \bibinfo{pages}{12--18}.
\newblock


\bibitem[Chajewska et~al\mbox{.}(2000)]%
        {chajewska2000making}
\bibfield{author}{\bibinfo{person}{Urszula Chajewska}, \bibinfo{person}{Daphne
  Koller}, {and} \bibinfo{person}{Ronald Parr}.}
  \bibinfo{year}{2000}\natexlab{}.
\newblock \showarticletitle{Making rational decisions using adaptive utility
  elicitation}. In \bibinfo{booktitle}{\emph{AAAI Proceedings}}.
  \bibinfo{publisher}{AAAI}, \bibinfo{address}{New York, NY},
  \bibinfo{pages}{363--369}.
\newblock


\bibitem[Charmaz(2006)]%
        {charmaz2006constructing}
\bibfield{author}{\bibinfo{person}{Kathy Charmaz}.}
  \bibinfo{year}{2006}\natexlab{}.
\newblock \bibinfo{booktitle}{\emph{Constructing grounded theory: A practical
  guide through qualitative analysis}}.
\newblock \bibinfo{publisher}{Sage}, \bibinfo{address}{New York, NY}.
\newblock


\bibitem[Chen et~al\mbox{.}(2023)]%
        {chen2023diagnosing}
\bibfield{author}{\bibinfo{person}{Hao Chen}, \bibinfo{person}{Gonzalo~E
  Constante-Flores}, {and} \bibinfo{person}{Can Li}.}
  \bibinfo{year}{2023}\natexlab{}.
\newblock \showarticletitle{Diagnosing infeasible optimization problems using
  large language models}.
\newblock \bibinfo{journal}{\emph{arXiv preprint arXiv:2308.12923}}
  (\bibinfo{year}{2023}).
\newblock


\bibitem[Chen and Pu(2004)]%
        {chen2004survey}
\bibfield{author}{\bibinfo{person}{Li Chen} {and} \bibinfo{person}{Pearl Pu}.}
  \bibinfo{year}{2004}\natexlab{}.
\newblock \bibinfo{booktitle}{\emph{Survey of preference elicitation methods}}.
\newblock \bibinfo{type}{{T}echnical {R}eport}. \bibinfo{institution}{EPFL}.
\newblock


\bibitem[Chen and Pu(2009)]%
        {chen2009interaction}
\bibfield{author}{\bibinfo{person}{Li Chen} {and} \bibinfo{person}{Pearl Pu}.}
  \bibinfo{year}{2009}\natexlab{}.
\newblock \showarticletitle{Interaction design guidelines on critiquing-based
  recommender systems}.
\newblock \bibinfo{journal}{\emph{User Modeling and User-Adapted Interaction}}
  \bibinfo{volume}{19} (\bibinfo{year}{2009}), \bibinfo{pages}{167--206}.
\newblock


\bibitem[Chen and Pu(2012)]%
        {chen2012critiquing}
\bibfield{author}{\bibinfo{person}{Li Chen} {and} \bibinfo{person}{Pearl Pu}.}
  \bibinfo{year}{2012}\natexlab{}.
\newblock \showarticletitle{Critiquing-based recommenders: Survey and emerging
  trends}.
\newblock \bibinfo{journal}{\emph{User Modeling and User-Adapted Interaction}}
  \bibinfo{volume}{22} (\bibinfo{year}{2012}), \bibinfo{pages}{125--150}.
\newblock


\bibitem[Chen et~al\mbox{.}(2021)]%
        {chen2021evaluating}
\bibfield{author}{\bibinfo{person}{Mark Chen}, \bibinfo{person}{Jerry Tworek},
  \bibinfo{person}{Heewoo Jun}, \bibinfo{person}{Qiming Yuan},
  \bibinfo{person}{Henrique Ponde De~Oliveira Pinto}, \bibinfo{person}{Jared
  Kaplan}, \bibinfo{person}{Harri Edwards}, \bibinfo{person}{Yuri Burda},
  \bibinfo{person}{Nicholas Joseph}, \bibinfo{person}{Greg Brockman},
  {et~al\mbox{.}}} \bibinfo{year}{2021}\natexlab{}.
\newblock \showarticletitle{Evaluating large language models trained on code}.
\newblock \bibinfo{journal}{\emph{arXiv preprint arXiv:2107.03374}}
  (\bibinfo{year}{2021}).
\newblock


\bibitem[Cranshaw et~al\mbox{.}(2017)]%
        {cranshaw2017calendar}
\bibfield{author}{\bibinfo{person}{Justin Cranshaw}, \bibinfo{person}{Emad
  Elwany}, \bibinfo{person}{Todd Newman}, \bibinfo{person}{Rafal Kocielnik},
  \bibinfo{person}{Bowen Yu}, \bibinfo{person}{Sandeep Soni},
  \bibinfo{person}{Jaime Teevan}, {and} \bibinfo{person}{Andr{\'e}s
  Monroy-Hern{\'a}ndez}.} \bibinfo{year}{2017}\natexlab{}.
\newblock \showarticletitle{Calendar.help: Designing a workflow-based
  scheduling agent with humans in the loop}. In
  \bibinfo{booktitle}{\emph{Proceedings of the 2017 CHI Conference on Human
  Factors in Computing Systems}}. \bibinfo{pages}{2382--2393}.
\newblock


\bibitem[Curhan et~al\mbox{.}(2004)]%
        {curhan2004dynamic}
\bibfield{author}{\bibinfo{person}{Jared~R Curhan}, \bibinfo{person}{Margaret~A
  Neale}, {and} \bibinfo{person}{Lee Ross}.} \bibinfo{year}{2004}\natexlab{}.
\newblock \showarticletitle{Dynamic valuation: Preference changes in the
  context of face-to-face negotiation}.
\newblock \bibinfo{journal}{\emph{Journal of Experimental Social Psychology}}
  \bibinfo{volume}{40}, \bibinfo{number}{2} (\bibinfo{year}{2004}),
  \bibinfo{pages}{142--151}.
\newblock


\bibitem[Dakle et~al\mbox{.}(2023)]%
        {dakle2023ner4opt}
\bibfield{author}{\bibinfo{person}{Parag~Pravin Dakle}, \bibinfo{person}{Serdar
  Kad{\i}o{\u{g}}lu}, \bibinfo{person}{Karthik Uppuluri},
  \bibinfo{person}{Regina Politi}, \bibinfo{person}{Preethi Raghavan},
  \bibinfo{person}{SaiKrishna Rallabandi}, {and} \bibinfo{person}{Ravisutha
  Srinivasamurthy}.} \bibinfo{year}{2023}\natexlab{}.
\newblock \showarticletitle{Ner4Opt: Named entity recognition for optimization
  modelling from natural language}. In \bibinfo{booktitle}{\emph{International
  Conference on Integration of Constraint Programming, Artificial Intelligence,
  and Operations Research}}. Springer, \bibinfo{pages}{299--319}.
\newblock


\bibitem[Dent et~al\mbox{.}(1992)]%
        {dent1992personal}
\bibfield{author}{\bibinfo{person}{Lisa Dent}, \bibinfo{person}{Jesus
  Boticario}, \bibinfo{person}{John McDermott}, \bibinfo{person}{Tom Mitchell},
  {and} \bibinfo{person}{David Zabowski}.} \bibinfo{year}{1992}\natexlab{}.
\newblock \showarticletitle{A personal learning apprentice}. In
  \bibinfo{booktitle}{\emph{Proceedings of the Tenth National Conference on
  Artificial Intelligence}}. \bibinfo{pages}{96--103}.
\newblock


\bibitem[Dibia(2023)]%
        {dibia2023lida}
\bibfield{author}{\bibinfo{person}{Victor Dibia}.}
  \bibinfo{year}{2023}\natexlab{}.
\newblock \showarticletitle{LIDA: A tool for automatic generation of
  grammar-agnostic visualizations and infographics using large language
  models}. In \bibinfo{booktitle}{\emph{Proceedings of the 61st Annual Meeting
  of the Association for Computational Linguistics (Volume 3: System
  Demonstrations)}}. \bibinfo{pages}{113--126}.
\newblock


\bibitem[Doyle(2004)]%
        {doyle2004prospects}
\bibfield{author}{\bibinfo{person}{Jon Doyle}.}
  \bibinfo{year}{2004}\natexlab{}.
\newblock \showarticletitle{Prospects for preferences}.
\newblock \bibinfo{journal}{\emph{Computational Intelligence}}
  \bibinfo{volume}{20}, \bibinfo{number}{2} (\bibinfo{year}{2004}),
  \bibinfo{pages}{111--136}.
\newblock


\bibitem[Edwards and Newman(1982)]%
        {edwards1992multiattribute}
\bibfield{author}{\bibinfo{person}{Ward Edwards} {and}
  \bibinfo{person}{J.~Robert Newman}.} \bibinfo{year}{1982}\natexlab{}.
\newblock \showarticletitle{Multiattribute evaluation}.
\newblock \bibinfo{journal}{\emph{Quantitative Applications in the Social
  Sciences}}  \bibinfo{volume}{26} (\bibinfo{year}{1982}),
  \bibinfo{pages}{7--32}.
\newblock


\bibitem[Gaver et~al\mbox{.}(1999)]%
        {gaver1999design}
\bibfield{author}{\bibinfo{person}{Bill Gaver}, \bibinfo{person}{Tony Dunne},
  {and} \bibinfo{person}{Elena Pacenti}.} \bibinfo{year}{1999}\natexlab{}.
\newblock \showarticletitle{Design: Cultural probes}.
\newblock \bibinfo{journal}{\emph{Interactions}} \bibinfo{volume}{6},
  \bibinfo{number}{1} (\bibinfo{year}{1999}), \bibinfo{pages}{21--29}.
\newblock


\bibitem[Gervasio et~al\mbox{.}(2005)]%
        {gervasio2005active}
\bibfield{author}{\bibinfo{person}{Melinda~T Gervasio},
  \bibinfo{person}{Michael~D Moffitt}, \bibinfo{person}{Martha~E Pollack},
  \bibinfo{person}{Joseph~M Taylor}, {and} \bibinfo{person}{Tomas~E Uribe}.}
  \bibinfo{year}{2005}\natexlab{}.
\newblock \showarticletitle{Active preference learning for personalized
  calendar scheduling assistance}. In \bibinfo{booktitle}{\emph{Proceedings of
  the 10th International Conference on Intelligent User Interfaces}}.
  \bibinfo{pages}{90--97}.
\newblock


\bibitem[Goldsmith and Junker(2008)]%
        {goldsmith2008preference}
\bibfield{author}{\bibinfo{person}{Judy Goldsmith} {and}
  \bibinfo{person}{Ulrich Junker}.} \bibinfo{year}{2008}\natexlab{}.
\newblock \showarticletitle{Preference handling for artificial intelligence}.
\newblock \bibinfo{journal}{\emph{AI Magazine}} \bibinfo{volume}{29},
  \bibinfo{number}{4} (\bibinfo{year}{2008}), \bibinfo{pages}{9}.
\newblock


\bibitem[Graham and Rouncefield(2008)]%
        {graham2008probes}
\bibfield{author}{\bibinfo{person}{Connor Graham} {and} \bibinfo{person}{Mark
  Rouncefield}.} \bibinfo{year}{2008}\natexlab{}.
\newblock \showarticletitle{Probes and participation}. In
  \bibinfo{booktitle}{\emph{Proceedings of the Tenth Anniversary Conference on
  Participatory Design 2008}}. \bibinfo{publisher}{ACM}, \bibinfo{address}{New
  York, NY}, \bibinfo{pages}{194--197}.
\newblock


\bibitem[Gregory et~al\mbox{.}(1993)]%
        {gregory1993valuing}
\bibfield{author}{\bibinfo{person}{Robin Gregory}, \bibinfo{person}{Sarah
  Lichtenstein}, {and} \bibinfo{person}{Paul Slovic}.}
  \bibinfo{year}{1993}\natexlab{}.
\newblock \showarticletitle{Valuing environmental resources: A constructive
  approach}.
\newblock \bibinfo{journal}{\emph{Journal of Risk and Uncertainty}}
  \bibinfo{volume}{7}, \bibinfo{number}{2} (\bibinfo{year}{1993}),
  \bibinfo{pages}{177--197}.
\newblock


\bibitem[Hammond et~al\mbox{.}(2015)]%
        {hammond2015smart}
\bibfield{author}{\bibinfo{person}{John~S Hammond}, \bibinfo{person}{Ralph~L
  Keeney}, {and} \bibinfo{person}{Howard Raiffa}.}
  \bibinfo{year}{2015}\natexlab{}.
\newblock \bibinfo{booktitle}{\emph{Smart choices: A practical guide to making
  better decisions}}.
\newblock \bibinfo{publisher}{Harvard Business Review Press},
  \bibinfo{address}{Cambridge, MA}.
\newblock


\bibitem[Haynes et~al\mbox{.}(1997)]%
        {haynes1997automated}
\bibfield{author}{\bibinfo{person}{Thomas Haynes}, \bibinfo{person}{Sandip
  Sen}, \bibinfo{person}{Neeraj Arora}, {and} \bibinfo{person}{Rajani
  Nadella}.} \bibinfo{year}{1997}\natexlab{}.
\newblock \showarticletitle{An automated meeting scheduling system that
  utilizes user preferences}. In \bibinfo{booktitle}{\emph{Proceedings of the
  First International Conference on Autonomous Agents}}.
  \bibinfo{publisher}{ACM}, \bibinfo{address}{New York, NY},
  \bibinfo{pages}{308--315}.
\newblock


\bibitem[Holstein et~al\mbox{.}(2023)]%
        {holstein2023toward}
\bibfield{author}{\bibinfo{person}{Kenneth Holstein}, \bibinfo{person}{Maria
  De-Arteaga}, \bibinfo{person}{Lakshmi Tumati}, {and}
  \bibinfo{person}{Yanghuidi Cheng}.} \bibinfo{year}{2023}\natexlab{}.
\newblock \showarticletitle{Toward supporting perceptual complementarity in
  human-AI collaboration via reflection on unobservables}.
\newblock \bibinfo{journal}{\emph{Proceedings of the ACM on Human-Computer
  Interaction}} \bibinfo{volume}{7}, \bibinfo{number}{CSCW1}
  (\bibinfo{year}{2023}), \bibinfo{pages}{1--20}.
\newblock


\bibitem[Hutchinson et~al\mbox{.}(2003)]%
        {hutchinson2003technology}
\bibfield{author}{\bibinfo{person}{Hilary Hutchinson}, \bibinfo{person}{Wendy
  Mackay}, \bibinfo{person}{Bo Westerlund}, \bibinfo{person}{Benjamin~B
  Bederson}, \bibinfo{person}{Allison Druin}, \bibinfo{person}{Catherine
  Plaisant}, \bibinfo{person}{Michel Beaudouin-Lafon},
  \bibinfo{person}{St{\'e}phane Conversy}, \bibinfo{person}{Helen Evans},
  \bibinfo{person}{Heiko Hansen}, {et~al\mbox{.}}}
  \bibinfo{year}{2003}\natexlab{}.
\newblock \showarticletitle{Technology probes: Inspiring design for and with
  families}. In \bibinfo{booktitle}{\emph{Proceedings of the SIGCHI Conference
  on Human Factors in Computing Systems}}. \bibinfo{pages}{17--24}.
\newblock


\bibitem[Janis and Mann(1977)]%
        {janis1977decision}
\bibfield{author}{\bibinfo{person}{Irving~L Janis} {and} \bibinfo{person}{Leon
  Mann}.} \bibinfo{year}{1977}\natexlab{}.
\newblock \bibinfo{booktitle}{\emph{Decision making: A psychological analysis
  of conflict, choice, and commitment}}.
\newblock \bibinfo{publisher}{Free Press}, \bibinfo{address}{New York, NY}.
\newblock


\bibitem[Jannach et~al\mbox{.}(2021)]%
        {jannach2021survey}
\bibfield{author}{\bibinfo{person}{Dietmar Jannach}, \bibinfo{person}{Ahtsham
  Manzoor}, \bibinfo{person}{Wanling Cai}, {and} \bibinfo{person}{Li Chen}.}
  \bibinfo{year}{2021}\natexlab{}.
\newblock \showarticletitle{A survey on conversational recommender systems}.
\newblock \bibinfo{journal}{\emph{ACM Computing Surveys (CSUR)}}
  \bibinfo{volume}{54}, \bibinfo{number}{5} (\bibinfo{year}{2021}),
  \bibinfo{pages}{1--36}.
\newblock


\bibitem[Johnson et~al\mbox{.}(2005)]%
        {johnson2005making}
\bibfield{author}{\bibinfo{person}{Eric~J Johnson}, \bibinfo{person}{Mary
  Steffel}, {and} \bibinfo{person}{Daniel~G Goldstein}.}
  \bibinfo{year}{2005}\natexlab{}.
\newblock \showarticletitle{Making better decisions: From measuring to
  constructing preferences}.
\newblock \bibinfo{journal}{\emph{Health Psychology}} \bibinfo{volume}{24},
  \bibinfo{number}{4S} (\bibinfo{year}{2005}), \bibinfo{pages}{S17}.
\newblock


\bibitem[Keeney and Raiffa(1993)]%
        {keeney1993decisions}
\bibfield{author}{\bibinfo{person}{Ralph~L Keeney} {and}
  \bibinfo{person}{Howard Raiffa}.} \bibinfo{year}{1993}\natexlab{}.
\newblock \bibinfo{booktitle}{\emph{Decisions with multiple objectives:
  Preferences and value trade-offs}}.
\newblock \bibinfo{publisher}{Cambridge University Press},
  \bibinfo{address}{Cambridge, UK}.
\newblock


\bibitem[Kim et~al\mbox{.}(2018)]%
        {kim2018learning}
\bibfield{author}{\bibinfo{person}{Donghyeon Kim}, \bibinfo{person}{Jinhyuk
  Lee}, \bibinfo{person}{Donghee Choi}, \bibinfo{person}{Jaehoon Choi}, {and}
  \bibinfo{person}{Jaewoo Kang}.} \bibinfo{year}{2018}\natexlab{}.
\newblock \showarticletitle{Learning user preferences and understanding
  calendar contexts for event scheduling}. In
  \bibinfo{booktitle}{\emph{Proceedings of the 27th ACM International
  Conference on Information and Knowledge Management}}.
  \bibinfo{publisher}{ACM}, \bibinfo{address}{New York, NY},
  \bibinfo{pages}{337--346}.
\newblock


\bibitem[Krzywicki et~al\mbox{.}(2010)]%
        {krzywicki2010adaptive}
\bibfield{author}{\bibinfo{person}{Alfred Krzywicki}, \bibinfo{person}{Wayne
  Wobcke}, {and} \bibinfo{person}{Anna Wong}.} \bibinfo{year}{2010}\natexlab{}.
\newblock \showarticletitle{An adaptive calendar assistant using pattern mining
  for user preference modelling}. In \bibinfo{booktitle}{\emph{Proceedings of
  the 15th International Conference on Intelligent User Interfaces}}.
  \bibinfo{publisher}{ACM}, \bibinfo{address}{New York, NY},
  \bibinfo{pages}{71--80}.
\newblock


\bibitem[Langer et~al\mbox{.}(2021)]%
        {langer2021we}
\bibfield{author}{\bibinfo{person}{Markus Langer}, \bibinfo{person}{Daniel
  Oster}, \bibinfo{person}{Timo Speith}, \bibinfo{person}{Holger Hermanns},
  \bibinfo{person}{Lena K{\"a}stner}, \bibinfo{person}{Eva Schmidt},
  \bibinfo{person}{Andreas Sesing}, {and} \bibinfo{person}{Kevin Baum}.}
  \bibinfo{year}{2021}\natexlab{}.
\newblock \showarticletitle{What do we want from explainable artificial
  intelligence (XAI)? A stakeholder perspective on XAI and a conceptual model
  guiding interdisciplinary XAI research}.
\newblock \bibinfo{journal}{\emph{Artificial Intelligence}}
  \bibinfo{volume}{296} (\bibinfo{year}{2021}), \bibinfo{pages}{103473}.
\newblock


\bibitem[Lazar(2024)]%
        {lazar2022legitimacy}
\bibfield{author}{\bibinfo{person}{Seth Lazar}.}
  \bibinfo{year}{2024}\natexlab{}.
\newblock \showarticletitle{Legitimacy, authority, and democratic duties of
  explanation}.
\newblock \bibinfo{journal}{\emph{Oxford Studies in Political Philosophy}}
  \bibinfo{volume}{10} (\bibinfo{year}{2024}), \bibinfo{pages}{28}.
\newblock


\bibitem[Lebovitz et~al\mbox{.}(2022)]%
        {lebovitz2022engage}
\bibfield{author}{\bibinfo{person}{Sarah Lebovitz}, \bibinfo{person}{Hila
  Lifshitz-Assaf}, {and} \bibinfo{person}{Natalia Levina}.}
  \bibinfo{year}{2022}\natexlab{}.
\newblock \showarticletitle{To engage or not to engage with AI for critical
  judgments: How professionals deal with opacity when using AI for medical
  diagnosis}.
\newblock \bibinfo{journal}{\emph{Organization Science}} \bibinfo{volume}{33},
  \bibinfo{number}{1} (\bibinfo{year}{2022}), \bibinfo{pages}{126--148}.
\newblock


\bibitem[Lee and See(2004)]%
        {lee2004trust}
\bibfield{author}{\bibinfo{person}{John~D Lee} {and} \bibinfo{person}{Katrina~A
  See}.} \bibinfo{year}{2004}\natexlab{}.
\newblock \showarticletitle{Trust in automation: Designing for appropriate
  reliance}.
\newblock \bibinfo{journal}{\emph{Human Factors}} \bibinfo{volume}{46},
  \bibinfo{number}{1} (\bibinfo{year}{2004}), \bibinfo{pages}{50--80}.
\newblock


\bibitem[Li et~al\mbox{.}(2023a)]%
        {li2023large}
\bibfield{author}{\bibinfo{person}{Beibin Li}, \bibinfo{person}{Konstantina
  Mellou}, \bibinfo{person}{Bo Zhang}, \bibinfo{person}{Jeevan Pathuri}, {and}
  \bibinfo{person}{Ishai Menache}.} \bibinfo{year}{2023}\natexlab{a}.
\newblock \showarticletitle{Large language models for supply chain
  optimization}.
\newblock \bibinfo{journal}{\emph{arXiv preprint arXiv:2307.03875}}
  (\bibinfo{year}{2023}).
\newblock


\bibitem[Li et~al\mbox{.}(2023b)]%
        {li2023eliciting}
\bibfield{author}{\bibinfo{person}{Belinda~Z Li}, \bibinfo{person}{Alex
  Tamkin}, \bibinfo{person}{Noah Goodman}, {and} \bibinfo{person}{Jacob
  Andreas}.} \bibinfo{year}{2023}\natexlab{b}.
\newblock \showarticletitle{Eliciting human preferences with language models}.
\newblock \bibinfo{journal}{\emph{arXiv preprint arXiv:2310.11589}}
  (\bibinfo{year}{2023}).
\newblock


\bibitem[Lichtenstein and Slovic(2006)]%
        {lichtenstein2006construction}
\bibfield{author}{\bibinfo{person}{Sarah Lichtenstein} {and}
  \bibinfo{person}{Paul Slovic}.} \bibinfo{year}{2006}\natexlab{}.
\newblock \bibinfo{booktitle}{\emph{The construction of preference}}.
\newblock \bibinfo{publisher}{Cambridge University Press}.
\newblock


\bibitem[Lin et~al\mbox{.}(2024)]%
        {lin2023decision}
\bibfield{author}{\bibinfo{person}{Jessy Lin}, \bibinfo{person}{Nicholas
  Tomlin}, \bibinfo{person}{Jacob Andreas}, {and} \bibinfo{person}{Jason
  Eisner}.} \bibinfo{year}{2024}\natexlab{}.
\newblock \showarticletitle{Decision-Oriented Dialogue for Human-AI
  Collaboration}. In \bibinfo{booktitle}{\emph{ICLR 2024 Workshop on Large
  Language Model (LLM) Agents}}.
\newblock


\bibitem[McGinty and Reilly(2010)]%
        {mcginty2010evolution}
\bibfield{author}{\bibinfo{person}{Lorraine McGinty} {and}
  \bibinfo{person}{James Reilly}.} \bibinfo{year}{2010}\natexlab{}.
\newblock \showarticletitle{On the evolution of critiquing recommenders}.
\newblock In \bibinfo{booktitle}{\emph{Recommender systems handbook}}.
  \bibinfo{publisher}{Springer}, \bibinfo{address}{New York, NY},
  \bibinfo{pages}{419--453}.
\newblock


\bibitem[Mialon et~al\mbox{.}(2023)]%
        {mialon2023augmented}
\bibfield{author}{\bibinfo{person}{Gr{\'e}goire Mialon},
  \bibinfo{person}{Roberto Dessi}, \bibinfo{person}{Maria Lomeli},
  \bibinfo{person}{Christoforos Nalmpantis}, \bibinfo{person}{Ramakanth
  Pasunuru}, \bibinfo{person}{Roberta Raileanu}, \bibinfo{person}{Baptiste
  Roziere}, \bibinfo{person}{Timo Schick}, \bibinfo{person}{Jane Dwivedi-Yu},
  \bibinfo{person}{Asli Celikyilmaz}, \bibinfo{person}{Edouard Grave},
  \bibinfo{person}{Yann LeCun}, {and} \bibinfo{person}{Thomas Scialom}.}
  \bibinfo{year}{2023}\natexlab{}.
\newblock \showarticletitle{Augmented Language Models: a Survey}.
\newblock \bibinfo{journal}{\emph{Transactions on Machine Learning Research}}
  (\bibinfo{year}{2023}).
\newblock
\showISSN{2835-8856}
\urldef\tempurl%
\url{https://openreview.net/forum?id=jh7wH2AzKK}
\showURL{%
\tempurl}
\newblock
\shownote{Survey Certification}.


\bibitem[Miller(2019)]%
        {miller2019explanation}
\bibfield{author}{\bibinfo{person}{Tim Miller}.}
  \bibinfo{year}{2019}\natexlab{}.
\newblock \showarticletitle{Explanation in artificial intelligence: Insights
  from the social sciences}.
\newblock \bibinfo{journal}{\emph{Artificial intelligence}}
  \bibinfo{volume}{267} (\bibinfo{year}{2019}), \bibinfo{pages}{1--38}.
\newblock


\bibitem[Mills and O’Neal(1971)]%
        {mills1971anticipated}
\bibfield{author}{\bibinfo{person}{Judson Mills} {and} \bibinfo{person}{Edgar
  O’Neal}.} \bibinfo{year}{1971}\natexlab{}.
\newblock \showarticletitle{Anticipated choice, attention, and halo effect}.
\newblock \bibinfo{journal}{\emph{Psychonomic Science}} \bibinfo{volume}{22},
  \bibinfo{number}{4} (\bibinfo{year}{1971}), \bibinfo{pages}{231--233}.
\newblock


\bibitem[Mitchell et~al\mbox{.}(1994)]%
        {mitchell1994experience}
\bibfield{author}{\bibinfo{person}{Tom~M Mitchell}, \bibinfo{person}{Rich
  Caruana}, \bibinfo{person}{Dayne Freitag}, \bibinfo{person}{John McDermott},
  \bibinfo{person}{David Zabowski}, {et~al\mbox{.}}}
  \bibinfo{year}{1994}\natexlab{}.
\newblock \showarticletitle{Experience with a learning personal assistant}.
\newblock \bibinfo{journal}{\emph{Commun. ACM}} \bibinfo{volume}{37},
  \bibinfo{number}{7} (\bibinfo{year}{1994}), \bibinfo{pages}{80--91}.
\newblock


\bibitem[Mok et~al\mbox{.}(2023)]%
        {mok2023challenging}
\bibfield{author}{\bibinfo{person}{Lillio Mok}, \bibinfo{person}{Lu Sun},
  \bibinfo{person}{Shilad Sen}, {and} \bibinfo{person}{Bahareh Sarrafzadeh}.}
  \bibinfo{year}{2023}\natexlab{}.
\newblock \showarticletitle{Challenging but connective: Large-scale
  characteristics of synchronous collaboration across time zones}. In
  \bibinfo{booktitle}{\emph{Proceedings of the 2023 CHI Conference on Human
  Factors in Computing Systems}}. \bibinfo{publisher}{ACM},
  \bibinfo{address}{New York, NY}, \bibinfo{pages}{1--17}.
\newblock


\bibitem[Montgomery and Will{\'e}n(2007)]%
        {montgomery1999decision}
\bibfield{author}{\bibinfo{person}{Henry Montgomery} {and}
  \bibinfo{person}{Helena Will{\'e}n}.} \bibinfo{year}{2007}\natexlab{}.
\newblock \showarticletitle{Decision making and action: The search for a good
  structure}.
\newblock In \bibinfo{booktitle}{\emph{Judgment and decision making}}.
  \bibinfo{publisher}{Psychology Press}, \bibinfo{pages}{147--173}.
\newblock


\bibitem[Nakano et~al\mbox{.}(2021)]%
        {nakano2021webgpt}
\bibfield{author}{\bibinfo{person}{Reiichiro Nakano}, \bibinfo{person}{Jacob
  Hilton}, \bibinfo{person}{Suchir Balaji}, \bibinfo{person}{Jeff Wu},
  \bibinfo{person}{Long Ouyang}, \bibinfo{person}{Christina Kim},
  \bibinfo{person}{Christopher Hesse}, \bibinfo{person}{Shantanu Jain},
  \bibinfo{person}{Vineet Kosaraju}, \bibinfo{person}{William Saunders},
  {et~al\mbox{.}}} \bibinfo{year}{2021}\natexlab{}.
\newblock \showarticletitle{WebGPT: Browser-assisted question-answering with
  human feedback}.
\newblock \bibinfo{journal}{\emph{arXiv preprint arXiv:2112.09332}}
  (\bibinfo{year}{2021}).
\newblock


\bibitem[Ning et~al\mbox{.}(2023)]%
        {ning2023novel}
\bibfield{author}{\bibinfo{person}{Yuting Ning}, \bibinfo{person}{Jiayu Liu},
  \bibinfo{person}{Longhu Qin}, \bibinfo{person}{Tong Xiao},
  \bibinfo{person}{Shangzi Xue}, \bibinfo{person}{Zhenya Huang},
  \bibinfo{person}{Qi Liu}, \bibinfo{person}{Enhong Chen}, {and}
  \bibinfo{person}{Jinze Wu}.} \bibinfo{year}{2023}\natexlab{}.
\newblock \showarticletitle{A novel approach for auto-formulation of
  optimization problems}.
\newblock \bibinfo{journal}{\emph{arXiv preprint arXiv:2302.04643}}
  (\bibinfo{year}{2023}).
\newblock


\bibitem[Oh and Smith(2005)]%
        {oh2005calendar}
\bibfield{author}{\bibinfo{person}{Jean Oh} {and} \bibinfo{person}{Stephen~F
  Smith}.} \bibinfo{year}{2005}\natexlab{}.
\newblock \showarticletitle{Calendar assistants that learn preferences}. In
  \bibinfo{booktitle}{\emph{AAAI Spring Symposium: Persistent Assistants:
  Living and Working with AI}}. \bibinfo{pages}{7--13}.
\newblock


\bibitem[Payne et~al\mbox{.}(1992)]%
        {payne1992behavioral}
\bibfield{author}{\bibinfo{person}{John~W Payne}, \bibinfo{person}{James~R
  Bettman}, {and} \bibinfo{person}{Eric~J Johnson}.}
  \bibinfo{year}{1992}\natexlab{}.
\newblock \showarticletitle{Behavioral decision research: A constructive
  processing perspective}.
\newblock \bibinfo{journal}{\emph{Annual Review of Psychology}}
  \bibinfo{volume}{43}, \bibinfo{number}{1} (\bibinfo{year}{1992}),
  \bibinfo{pages}{87--131}.
\newblock


\bibitem[Payne et~al\mbox{.}(1993)]%
        {payne1993adaptive}
\bibfield{author}{\bibinfo{person}{John~W Payne}, \bibinfo{person}{James~R
  Bettman}, {and} \bibinfo{person}{Eric~J Johnson}.}
  \bibinfo{year}{1993}\natexlab{}.
\newblock \bibinfo{booktitle}{\emph{The adaptive decision maker}}.
\newblock \bibinfo{publisher}{Cambridge University Press},
  \bibinfo{address}{Cambridge, UK}.
\newblock


\bibitem[Payne et~al\mbox{.}(2000)]%
        {payne2000measuring}
\bibfield{author}{\bibinfo{person}{John~W Payne}, \bibinfo{person}{James~R
  Bettman}, \bibinfo{person}{David~A Schkade}, \bibinfo{person}{Norbert
  Schwarz}, {and} \bibinfo{person}{Robin Gregory}.}
  \bibinfo{year}{2000}\natexlab{}.
\newblock \showarticletitle{Measuring constructed preferences: Towards a
  building code}.
\newblock \bibinfo{journal}{\emph{Elicitation of Preferences}}
  \bibinfo{volume}{19} (\bibinfo{year}{2000}), \bibinfo{pages}{243--275}.
\newblock


\bibitem[Peintner et~al\mbox{.}(2008)]%
        {peintner2008preferences}
\bibfield{author}{\bibinfo{person}{Bart Peintner}, \bibinfo{person}{Paolo
  Viappiani}, {and} \bibinfo{person}{Neil Yorke-Smith}.}
  \bibinfo{year}{2008}\natexlab{}.
\newblock \showarticletitle{Preferences in interactive systems: Technical
  challenges and case studies}.
\newblock \bibinfo{journal}{\emph{AI Magazine}} \bibinfo{volume}{29},
  \bibinfo{number}{4} (\bibinfo{year}{2008}), \bibinfo{pages}{13--13}.
\newblock


\bibitem[Pommeranz et~al\mbox{.}(2012)]%
        {pommeranz2012designing}
\bibfield{author}{\bibinfo{person}{Alina Pommeranz}, \bibinfo{person}{Joost
  Broekens}, \bibinfo{person}{Pascal Wiggers}, \bibinfo{person}{Willem-Paul
  Brinkman}, {and} \bibinfo{person}{Catholijn~M Jonker}.}
  \bibinfo{year}{2012}\natexlab{}.
\newblock \showarticletitle{Designing interfaces for explicit preference
  elicitation: A user-centered investigation of preference representation and
  elicitation process}.
\newblock \bibinfo{journal}{\emph{User Modeling and User-Adapted Interaction}}
  \bibinfo{volume}{22} (\bibinfo{year}{2012}), \bibinfo{pages}{357--397}.
\newblock


\bibitem[Prasath and Karande(2023)]%
        {prasath2023synthesis}
\bibfield{author}{\bibinfo{person}{Ganesh Prasath} {and}
  \bibinfo{person}{Shirish Karande}.} \bibinfo{year}{2023}\natexlab{}.
\newblock \showarticletitle{Synthesis of mathematical programs from natural
  language specifications}.
\newblock \bibinfo{journal}{\emph{arXiv preprint arXiv:2304.03287}}
  (\bibinfo{year}{2023}).
\newblock


\bibitem[Pu et~al\mbox{.}(2003)]%
        {pu2003user}
\bibfield{author}{\bibinfo{person}{Pearl Pu}, \bibinfo{person}{Boi Faltings},
  {and} \bibinfo{person}{Marc Torrens}.} \bibinfo{year}{2003}\natexlab{}.
\newblock \bibinfo{booktitle}{\emph{User-involved preference elicitation}}.
\newblock \bibinfo{type}{{T}echnical {R}eport}. \bibinfo{institution}{EPFL}.
\newblock


\bibitem[Qian et~al\mbox{.}(2023)]%
        {qian2022limitations}
\bibfield{author}{\bibinfo{person}{Jing Qian}, \bibinfo{person}{Hong Wang},
  \bibinfo{person}{Zekun Li}, \bibinfo{person}{Shiyang Li}, {and}
  \bibinfo{person}{Xifeng Yan}.} \bibinfo{year}{2023}\natexlab{}.
\newblock \showarticletitle{Limitations of language models in arithmetic and
  symbolic induction}. In \bibinfo{booktitle}{\emph{Proceedings of the 61st
  Annual Meeting of the Association for Computational Linguistics (Volume 1:
  Long Papers)}}. \bibinfo{pages}{9285--9298}.
\newblock


\bibitem[Ramamonjison et~al\mbox{.}(2022)]%
        {ramamonjison2022augmenting}
\bibfield{author}{\bibinfo{person}{Rindra Ramamonjison}, \bibinfo{person}{Haley
  Li}, \bibinfo{person}{Timothy Yu}, \bibinfo{person}{Shiqi He},
  \bibinfo{person}{Vishnu Rengan}, \bibinfo{person}{Amin Banitalebi-Dehkordi},
  \bibinfo{person}{Zirui Zhou}, {and} \bibinfo{person}{Yong Zhang}.}
  \bibinfo{year}{2022}\natexlab{}.
\newblock \showarticletitle{Augmenting operations research with
  auto-formulation of optimization models from problem descriptions}. In
  \bibinfo{booktitle}{\emph{Proceedings of the 2022 Conference on Empirical
  Methods in Natural Language Processing: Industry Track}}.
  \bibinfo{pages}{29--62}.
\newblock


\bibitem[Richter(1966)]%
        {richter1966revealed}
\bibfield{author}{\bibinfo{person}{Marcel~K Richter}.}
  \bibinfo{year}{1966}\natexlab{}.
\newblock \showarticletitle{Revealed preference theory}.
\newblock \bibinfo{journal}{\emph{Econometrica: Journal of the Econometric
  Society}}  \bibinfo{volume}{34} (\bibinfo{year}{1966}),
  \bibinfo{pages}{635--645}.
\newblock


\bibitem[Rossi(1999)]%
        {rossi1999constraint}
\bibfield{author}{\bibinfo{person}{Francesca Rossi}.}
  \bibinfo{year}{1999}\natexlab{}.
\newblock \showarticletitle{Constraint (logic) programming: A survey on
  research and applications}. In \bibinfo{booktitle}{\emph{Compulog Net/ERCIM
  Workshop on Constraints}}. Springer, \bibinfo{publisher}{Springer},
  \bibinfo{address}{New York, NY}, \bibinfo{pages}{40--74}.
\newblock


\bibitem[Russo et~al\mbox{.}(1996)]%
        {russo1996distortion}
\bibfield{author}{\bibinfo{person}{J~Edward Russo},
  \bibinfo{person}{Victoria~Husted Medvec}, {and} \bibinfo{person}{Margaret~G
  Meloy}.} \bibinfo{year}{1996}\natexlab{}.
\newblock \showarticletitle{The distortion of information during decisions}.
\newblock \bibinfo{journal}{\emph{Organizational Behavior and Human Decision
  Processes}} \bibinfo{volume}{66}, \bibinfo{number}{1} (\bibinfo{year}{1996}),
  \bibinfo{pages}{102--110}.
\newblock


\bibitem[Schemmer et~al\mbox{.}(2023)]%
        {schemmer2023appropriate}
\bibfield{author}{\bibinfo{person}{Max Schemmer}, \bibinfo{person}{Niklas
  Kuehl}, \bibinfo{person}{Carina Benz}, \bibinfo{person}{Andrea Bartos}, {and}
  \bibinfo{person}{Gerhard Satzger}.} \bibinfo{year}{2023}\natexlab{}.
\newblock \showarticletitle{Appropriate reliance on AI advice:
  Conceptualization and the effect of explanations}. In
  \bibinfo{booktitle}{\emph{Proceedings of the 28th International Conference on
  Intelligent User Interfaces}}. \bibinfo{pages}{410--422}.
\newblock


\bibitem[Schick et~al\mbox{.}(2024)]%
        {schick2023toolformer}
\bibfield{author}{\bibinfo{person}{Timo Schick}, \bibinfo{person}{Jane
  Dwivedi-Yu}, \bibinfo{person}{Roberto Dess{\`\i}}, \bibinfo{person}{Roberta
  Raileanu}, \bibinfo{person}{Maria Lomeli}, \bibinfo{person}{Eric Hambro},
  \bibinfo{person}{Luke Zettlemoyer}, \bibinfo{person}{Nicola Cancedda}, {and}
  \bibinfo{person}{Thomas Scialom}.} \bibinfo{year}{2024}\natexlab{}.
\newblock \showarticletitle{Toolformer: Language models can teach themselves to
  use tools}.
\newblock \bibinfo{journal}{\emph{Advances in Neural Information Processing
  Systems}}  \bibinfo{volume}{36} (\bibinfo{year}{2024}).
\newblock


\bibitem[Schoeffer et~al\mbox{.}(2024)]%
        {schoeffer2024explanations}
\bibfield{author}{\bibinfo{person}{Jakob Schoeffer}, \bibinfo{person}{Maria
  De-Arteaga}, {and} \bibinfo{person}{Niklas Kuehl}.}
  \bibinfo{year}{2024}\natexlab{}.
\newblock \showarticletitle{Explanations, fairness, and appropriate reliance in
  human-AI decision-making}. In \bibinfo{booktitle}{\emph{Proceedings of the
  CHI Conference on Human Factors in Computing Systems}}.
  \bibinfo{pages}{1--18}.
\newblock


\bibitem[Schoeffer et~al\mbox{.}(2023)]%
        {schoeffer2023interdependence}
\bibfield{author}{\bibinfo{person}{Jakob Schoeffer}, \bibinfo{person}{Johannes
  Jakubik}, \bibinfo{person}{Michael Voessing}, \bibinfo{person}{Niklas Kuehl},
  {and} \bibinfo{person}{Gerhard Satzger}.} \bibinfo{year}{2023}\natexlab{}.
\newblock \showarticletitle{On the interdependence of reliance behavior and
  accuracy in AI-assisted decision-making}.
\newblock In \bibinfo{booktitle}{\emph{HHAI 2023: Augmenting Human Intellect}}.
  \bibinfo{publisher}{IOS Press}, \bibinfo{pages}{46--59}.
\newblock


\bibitem[Sen et~al\mbox{.}(1997)]%
        {sen1997satisfying}
\bibfield{author}{\bibinfo{person}{Sandip Sen}, \bibinfo{person}{Thomas
  Haynes}, {and} \bibinfo{person}{Neeraj Arora}.}
  \bibinfo{year}{1997}\natexlab{}.
\newblock \showarticletitle{Satisfying user preferences while negotiating
  meetings}.
\newblock \bibinfo{journal}{\emph{International Journal of Human-Computer
  Studies}} \bibinfo{volume}{47}, \bibinfo{number}{3} (\bibinfo{year}{1997}),
  \bibinfo{pages}{407--427}.
\newblock


\bibitem[Simon et~al\mbox{.}(2008)]%
        {simon2008transience}
\bibfield{author}{\bibinfo{person}{Dan Simon}, \bibinfo{person}{Daniel~C
  Krawczyk}, \bibinfo{person}{Airom Bleicher}, {and} \bibinfo{person}{Keith~J
  Holyoak}.} \bibinfo{year}{2008}\natexlab{}.
\newblock \showarticletitle{The transience of constructed preferences}.
\newblock \bibinfo{journal}{\emph{Journal of Behavioral Decision Making}}
  \bibinfo{volume}{21}, \bibinfo{number}{1} (\bibinfo{year}{2008}),
  \bibinfo{pages}{1--14}.
\newblock


\bibitem[Slovic(1995)]%
        {Slovic1995construction}
\bibfield{author}{\bibinfo{person}{Paul Slovic}.}
  \bibinfo{year}{1995}\natexlab{}.
\newblock \showarticletitle{The construction of preference}.
\newblock \bibinfo{journal}{\emph{American Psychologist}}  \bibinfo{volume}{50}
  (\bibinfo{year}{1995}), \bibinfo{pages}{364--371}.
\newblock


\bibitem[Sun et~al\mbox{.}(2024)]%
        {baharcscw}
\bibfield{author}{\bibinfo{person}{Lu Sun}, \bibinfo{person}{Lillio Mok},
  \bibinfo{person}{Shilad Sen}, {and} \bibinfo{person}{Bahar Sarrafzadeh}.}
  \bibinfo{year}{2024}\natexlab{}.
\newblock \showarticletitle{Rhythm of work: Mixed-methods characterization of
  information workers scheduling preferences and practices}.
\newblock \bibinfo{journal}{\emph{The 27th ACM Conference on Computer-Supported
  Cooperative Work and Social Computing}} (\bibinfo{year}{2024}).
\newblock


\bibitem[Tabakhi(2017)]%
        {tabakhi2017preference}
\bibfield{author}{\bibinfo{person}{Atena~M Tabakhi}.}
  \bibinfo{year}{2017}\natexlab{}.
\newblock \showarticletitle{Preference elicitation in DCOPs for scheduling
  devices in smart buildings}. In \bibinfo{booktitle}{\emph{Proceedings of the
  AAAI Conference on Artificial Intelligence}}. \bibinfo{publisher}{AAAI},
  \bibinfo{address}{New York, NY}, \bibinfo{pages}{64--78}.
\newblock


\bibitem[Tabakhi(2021)]%
        {tabakhi2021preference}
\bibfield{author}{\bibinfo{person}{Atena~M Tabakhi}.}
  \bibinfo{year}{2021}\natexlab{}.
\newblock \emph{\bibinfo{title}{Preference elicitation in constraint-based
  models: Models, algorithms, and applications}}.
\newblock \bibinfo{thesistype}{Ph.\,D. Dissertation}.
  \bibinfo{school}{Washington University in St. Louis}.
\newblock


\bibitem[Tabakhi et~al\mbox{.}(2017)]%
        {tabakhi2017preference2}
\bibfield{author}{\bibinfo{person}{Atena~M Tabakhi}, \bibinfo{person}{Tiep Le},
  \bibinfo{person}{Ferdinando Fioretto}, {and} \bibinfo{person}{William Yeoh}.}
  \bibinfo{year}{2017}\natexlab{}.
\newblock \showarticletitle{Preference elicitation for DCOPs}. In
  \bibinfo{booktitle}{\emph{Principles and Practice of Constraint
  Programming}}. \bibinfo{publisher}{Springer}, \bibinfo{address}{New York,
  NY}, \bibinfo{pages}{278--296}.
\newblock


\bibitem[Tabakhi et~al\mbox{.}(2022)]%
        {tabakhi2022incomplete}
\bibfield{author}{\bibinfo{person}{Atena~M Tabakhi}, \bibinfo{person}{William
  Yeoh}, {and} \bibinfo{person}{Roie Zivan}.} \bibinfo{year}{2022}\natexlab{}.
\newblock \showarticletitle{Incomplete distributed constraint optimization
  problems: Model, algorithms, and heuristics}. In
  \bibinfo{booktitle}{\emph{Distributed Artificial Intelligence}}.
  \bibinfo{publisher}{Springer}, \bibinfo{address}{New York, NY},
  \bibinfo{pages}{64--78}.
\newblock


\bibitem[Thoppilan et~al\mbox{.}(2022)]%
        {thoppilan2022lamda}
\bibfield{author}{\bibinfo{person}{Romal Thoppilan}, \bibinfo{person}{Daniel
  De~Freitas}, \bibinfo{person}{Jamie Hall}, \bibinfo{person}{Noam Shazeer},
  \bibinfo{person}{Apoorv Kulshreshtha}, \bibinfo{person}{Heng-Tze Cheng},
  \bibinfo{person}{Alicia Jin}, \bibinfo{person}{Taylor Bos},
  \bibinfo{person}{Leslie Baker}, \bibinfo{person}{Yu Du}, {et~al\mbox{.}}}
  \bibinfo{year}{2022}\natexlab{}.
\newblock \showarticletitle{LaMDA: Language models for dialog applications}.
\newblock \bibinfo{journal}{\emph{arXiv preprint arXiv:2201.08239}}
  (\bibinfo{year}{2022}).
\newblock


\bibitem[Tsouros et~al\mbox{.}(2023)]%
        {tsouros2023holy}
\bibfield{author}{\bibinfo{person}{Dimos Tsouros},
  \bibinfo{person}{H{\'e}l{\`e}ne Verhaeghe}, \bibinfo{person}{Serdar
  Kad{\i}o{\u{g}}lu}, {and} \bibinfo{person}{Tias Guns}.}
  \bibinfo{year}{2023}\natexlab{}.
\newblock \showarticletitle{Holy Grail 2.0: From natural language to constraint
  models}.
\newblock \bibinfo{journal}{\emph{arXiv preprint arXiv:2308.01589}}
  (\bibinfo{year}{2023}).
\newblock


\bibitem[Tullio et~al\mbox{.}(2002)]%
        {tullio2002augmenting}
\bibfield{author}{\bibinfo{person}{Joe Tullio}, \bibinfo{person}{Jeremy
  Goecks}, \bibinfo{person}{Elizabeth~D Mynatt}, {and} \bibinfo{person}{David~H
  Nguyen}.} \bibinfo{year}{2002}\natexlab{}.
\newblock \showarticletitle{Augmenting shared personal calendars}. In
  \bibinfo{booktitle}{\emph{Proceedings of the 15th Annual ACM Symposium on
  User Interface Software and Technology}}. \bibinfo{publisher}{ACM},
  \bibinfo{address}{New York, NY}, \bibinfo{pages}{11--20}.
\newblock


\bibitem[Viappiani et~al\mbox{.}(2006)]%
        {viappiani2006evaluating}
\bibfield{author}{\bibinfo{person}{Paolo Viappiani}, \bibinfo{person}{Boi
  Faltings}, {and} \bibinfo{person}{Pearl Pu}.}
  \bibinfo{year}{2006}\natexlab{}.
\newblock \showarticletitle{Evaluating preference-based search tools: A tale of
  two approaches}. In \bibinfo{booktitle}{\emph{Proceedings of the Twenty-First
  National Conference on Artificial Intelligence (AAAI-06)}}. AAAI Press,
  \bibinfo{pages}{205--211}.
\newblock


\bibitem[Wei et~al\mbox{.}(2022)]%
        {wei2022chain}
\bibfield{author}{\bibinfo{person}{Jason Wei}, \bibinfo{person}{Xuezhi Wang},
  \bibinfo{person}{Dale Schuurmans}, \bibinfo{person}{Maarten Bosma},
  \bibinfo{person}{Fei Xia}, \bibinfo{person}{Ed Chi}, \bibinfo{person}{Quoc~V
  Le}, \bibinfo{person}{Denny Zhou}, {et~al\mbox{.}}}
  \bibinfo{year}{2022}\natexlab{}.
\newblock \showarticletitle{Chain-of-thought prompting elicits reasoning in
  large language models}.
\newblock \bibinfo{journal}{\emph{Advances in Neural Information Processing
  Systems}}  \bibinfo{volume}{35} (\bibinfo{year}{2022}),
  \bibinfo{pages}{24824--24837}.
\newblock


\bibitem[Welleck et~al\mbox{.}(2020)]%
        {welleck2019neural}
\bibfield{author}{\bibinfo{person}{Sean Welleck}, \bibinfo{person}{Ilia
  Kulikov}, \bibinfo{person}{Stephen Roller}, \bibinfo{person}{Emily Dinan},
  \bibinfo{person}{Kyunghyun Cho}, {and} \bibinfo{person}{Jason Weston}.}
  \bibinfo{year}{2020}\natexlab{}.
\newblock \showarticletitle{Neural text generation with unlikelihood training}.
  In \bibinfo{booktitle}{\emph{International Conference on Learning
  Representations}}.
\newblock


\bibitem[Xiao(2020)]%
        {xiao2020embedding}
\bibfield{author}{\bibinfo{person}{Yuanming Xiao}.}
  \bibinfo{year}{2020}\natexlab{}.
\newblock \emph{\bibinfo{title}{Embedding preference elicitation within the
  search for DCOP solutions}}.
\newblock \bibinfo{thesistype}{Ph.\,D. Dissertation}.
  \bibinfo{school}{Washington University in St. Louis}.
\newblock


\bibitem[Yao et~al\mbox{.}(2024)]%
        {yao2023tree}
\bibfield{author}{\bibinfo{person}{Shunyu Yao}, \bibinfo{person}{Dian Yu},
  \bibinfo{person}{Jeffrey Zhao}, \bibinfo{person}{Izhak Shafran},
  \bibinfo{person}{Tom Griffiths}, \bibinfo{person}{Yuan Cao}, {and}
  \bibinfo{person}{Karthik Narasimhan}.} \bibinfo{year}{2024}\natexlab{}.
\newblock \showarticletitle{Tree of thoughts: Deliberate problem solving with
  large language models}.
\newblock \bibinfo{journal}{\emph{Advances in Neural Information Processing
  Systems}}  \bibinfo{volume}{36} (\bibinfo{year}{2024}).
\newblock


\bibitem[Yao et~al\mbox{.}(2023)]%
        {yao2022react}
\bibfield{author}{\bibinfo{person}{Shunyu Yao}, \bibinfo{person}{Jeffrey Zhao},
  \bibinfo{person}{Dian Yu}, \bibinfo{person}{Nan Du}, \bibinfo{person}{Izhak
  Shafran}, \bibinfo{person}{Karthik~R Narasimhan}, {and} \bibinfo{person}{Yuan
  Cao}.} \bibinfo{year}{2023}\natexlab{}.
\newblock \showarticletitle{ReAct: Synergizing reasoning and acting in language
  models}. In \bibinfo{booktitle}{\emph{The Eleventh International Conference
  on Learning Representations}}.
\newblock


\bibitem[Yuan et~al\mbox{.}(2022)]%
        {yuan2022wordcraft}
\bibfield{author}{\bibinfo{person}{Ann Yuan}, \bibinfo{person}{Andy Coenen},
  \bibinfo{person}{Emily Reif}, {and} \bibinfo{person}{Daphne Ippolito}.}
  \bibinfo{year}{2022}\natexlab{}.
\newblock \showarticletitle{Wordcraft: Story writing with large language
  models}. In \bibinfo{booktitle}{\emph{27th International Conference on
  Intelligent User Interfaces}}. \bibinfo{publisher}{ACM},
  \bibinfo{address}{New York, NY}, \bibinfo{pages}{841--852}.
\newblock


\bibitem[Yuksekgonul et~al\mbox{.}(2024)]%
        {yuksekgonul2023attention}
\bibfield{author}{\bibinfo{person}{Mert Yuksekgonul}, \bibinfo{person}{Varun
  Chandrasekaran}, \bibinfo{person}{Erik Jones}, \bibinfo{person}{Suriya
  Gunasekar}, \bibinfo{person}{Ranjita Naik}, \bibinfo{person}{Hamid Palangi},
  \bibinfo{person}{Ece Kamar}, {and} \bibinfo{person}{Besmira Nushi}.}
  \bibinfo{year}{2024}\natexlab{}.
\newblock \showarticletitle{Attention satisfies: A constraint-satisfaction lens
  on factual errors of language models}. In \bibinfo{booktitle}{\emph{The
  Twelfth International Conference on Learning Representations}}.
\newblock


\end{thebibliography}

\appendix

\end{document}